\documentclass[prl,aps,twocolumn,showpacs,amsmath,superscriptaddress,amssymb]{revtex4-1}

\usepackage{bbm,dsfont}
\usepackage{graphicx}
\usepackage{psfrag}
\usepackage{epsfig}
\usepackage{color}
\usepackage[tight]{subfigure}
\definecolor{dred}{rgb}{0.7,0.0,0.0}
\usepackage{dcolumn}
\usepackage{bm}
\allowdisplaybreaks 
\usepackage{braket}
\usepackage{ulem} 
\usepackage{hyperref}
\hypersetup{colorlinks=true, linkcolor=blue, citecolor=blue, filecolor=blue, urlcolor=blue, pdftitle=, pdfauthor=, pdfsubject=, pdfkeywords=}

\newcommand{\bs}[1]{\ensuremath{\boldsymbol{#1}}}

\definecolor{orange}{rgb}{1,0.5,0}
\definecolor{black}{rgb}{0,0,0}

\begin{document}

\title{
A novel platform for two-dimensional chiral topological superconductivity
}

\date{\today}

\author{Jian Li} 
\affiliation{
Department of Physics,
Princeton University,
Princeton, NJ 08544, USA
            } 

\author{Titus Neupert} 
\affiliation{
Princeton Center for Theoretical Science,
Princeton University,
Princeton, NJ 08544, USA
            }

\author{Zhi Jun Wang}
\affiliation{
Department of Physics,
Princeton University,
Princeton, NJ 08544, USA
            } 

\author{A. H. MacDonald} 
\affiliation{
Department of Physics, University of Texas at Austin, Austin, TX 78712, USA            }

\author{A. Yazdani} 
\affiliation{
Department of Physics,
Princeton University,
Princeton, NJ 08544, USA
            } 

\author{B. Andrei Bernevig} 
\affiliation{
Department of Physics,
Princeton University,
Princeton, NJ 08544, USA
            } 

\begin{abstract}
We show that the surface of an $s$-wave superconductor decorated with a two-dimensional lattice of magnetic impurities can exhibit chiral topological superconductivity. If impurities order ferromagnetically and the superconducting surface supports a sufficiently strong Rashba-type spin-orbit coupling, Shiba sub-gap states at impurity locations can hybridize into Bogoliubov bands with non-vanishing, sometimes large, Chern number $C$. This topological superconductor supports $C$ chiral Majorana edge modes. 
We construct phase diagrams for model two-dimensional superconductors, accessing the dilute and dense magnetic impurity limits analytically and the intermediate regime numerically.
To address potential experimental systems, we identify stable configurations of 
ferromagnetic iron atoms on the Pb (111) surface and conclude that ferromagnetic adatoms on Pb surfaces can provide a versatile platform for two-dimensional topological superconductivity.
\end{abstract}

\maketitle

The chiral p-wave superconductor in two dimensions (2D) and the closely related fractional quantum Hall Pfaffian state at $\nu=5/2$ are the archetypal examples of topologically ordered states of matter that support non-Abelian anyonic excitations.~\cite{Moore91,Read2000,Ivanov01} The theoretical exploration of these states has shaped our understanding of topological order and is foundational for the distant goal of building a topological quantum computer.~\cite{Kitaev03,Nayak08,Tewari07} In contrast to the $\nu=5/2$ fractional quantum Hall state, however, to date chiral $p$-wave superconductivity has not been confirmed in any experimental system. The most prominent candidate system, superconducting Sr$_2$RuO$_4$,~\cite{Sarma06} has been the subject of an ongoing debate about its actual paring state.~\cite{Mackenzie03,Raghu10,Wang13,Scaffidi14}

Chiral superconductors break time-reversal symmetry (TRS). This hinders the formation of Cooper pairs, since orbital (and possibly paramagnetic) pair-breaking effects can come into play. Depairing is also the main hurdle for realizing a line of proposals in which layered heterostructures involving ferromagnets and $s$-wave superconductors are used to build an artificial 2D $p$-wave superconductor.~\cite{Sato09,Sau10,Alicea10} The guiding principle for these proposals is to design a band structure with a single normal-state Fermi surface with no spin-degeneracy. If Rashba spin-orbit coupling is present so that the states on this Fermi surface are not fully spin-polarized, and if $s$-wave superconductivity is proximity-induced in such a system, the effective pairing near the single Fermi surface is equivalent to that of a chiral $p$-wave superconductor.

\begin{figure}[htbp]
\begin{center}
\includegraphics[width=0.75\linewidth]{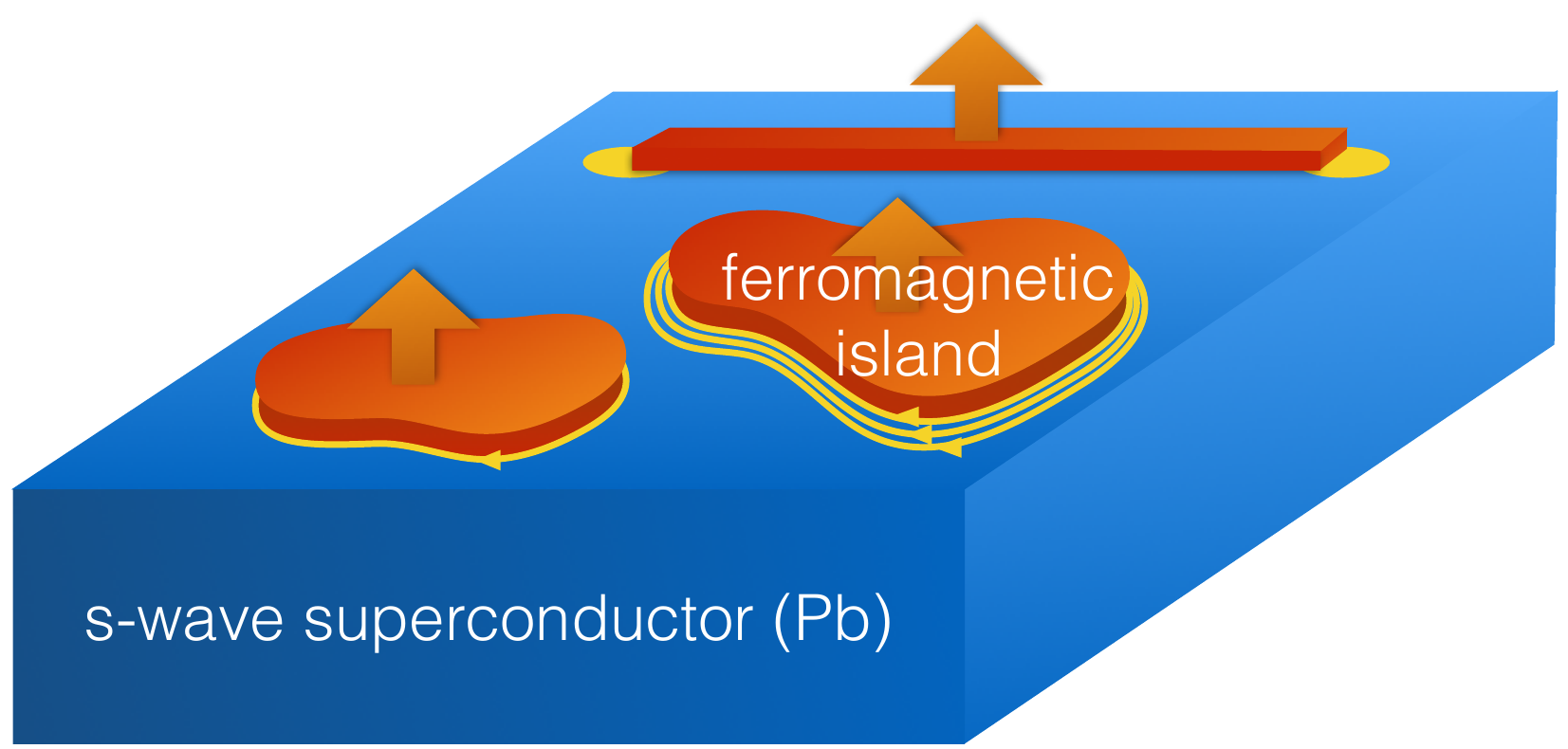}
\caption{
Concept of building a topological $p$-wave superconductor from magnetic impurity bound states
on the surface of an $s$-wave superconductor like Pb with strong spin-orbit coupling. A very thin layer of magnetic adatoms, such as Fe or Co, is deposited on the surface of Pb, and forms islands around which chiral Majorana edge modes are present (yellow). Depending on the density of the magnetic atoms, one or many Majorana modes may appear. 
A one-dimensional chain of  magnetic atoms may support localized Majorana modes at its end.
These modes can be detected by tunneling techniques. }
\label{fig: Setup}
\end{center}
\end{figure}

In one dimension (1D), based on the principle of combining spin-orbit coupling and externally applied magnetic fields, various groups have proposed engineering artificial realizations of $p$-wave superconductors.~\cite{Kitaev01,Lutchyn10,Oreg10,Potter10,Alicea11,Halperin12,Stanescu13}. An experimental realization of these proposals employing semiconductor nanowires with strong spin-orbit coupling~\cite{Mourik12} has reported Majorana fermion signatures. In this setup, the externally applied magnetic field has to be rather small (to avoid suppressing superconductivity). As the phase-space for the existence of a topological superconductor is controlled by the Zeeman gap \cite{Li14}, these systems require a delicate balance of the parameters involved (spin-orbit coupling, magnetic field and chemical potential) in order to create the topological superconductor. 

Recently, a 1D topological superconductor was realized in a system that is quite distinct but employs similar microscopic ingredients -- spin-orbit coupling, ferromagnetism, and s-wave superconductivity.~\cite{Nadj-Perge14}  A chain of magnetic Fe atoms is deposited on the surface of an $s$-wave superconductor with strong spin-orbit interactions. The Fe chain is ferromagnetically ordered~\cite{Nadj-Perge14} with a large magnetic moment, which takes the role of the magnetic field in the nanowire experiments. Unlike previous proposals, this ``magnetic field" is mostly localized on the Fe chain, with small leakage outside. Superconductivity is not destroyed along the chain. In this setup, the energy scale of the exchange coupling of the Fe atoms is typically much larger than that of the Rashba spin-orbit coupling and the superconducting pairing. 
The ferromagnetically ordered Fe atoms induce localized Shiba states within the gap of the superconductor.~\cite{Yu65,Shiba68,Rusinov68} The hybridization of these states forms the band structure of a 1D $p$-wave superconductor that supports Majorana end states.~\cite{Pientka13}  Because the Fe bands are fully spin split, no additional control over the chemical potential is necessary. A similar scenario applies when the Fe orbitals 
are magnetic but itinerant.~\cite{Li14}

In this Letter we point out that this strategy can also be successful in 2D.  
Magnetic adatoms on the surface of a superconductor with strong spin-orbit coupling, when arranged in a 2D lattice,~\cite{footnote1} can yield a 2D topological chiral $p$-wave superconductor whose chiral Majorana edge modes can be observed in Scanning Tunneling Microscope (STM) measurements (see Fig.~\ref{fig: Setup}).
To shed light on the rich range of possibilities, we analyze the topological properties of 
a system with dense local moments that are exchange coupled to a model 2D
superconductor, demonstrating that topological superconductors with higher Chern numbers,
and consequently multiple chiral Majorana edge channels, can easily occur.  We are also 
able to analyze the model's dilute magnetic impurity limit analytically and obtain numerical 
topological phase diagrams for intermediate impurity concentrations.
Based on density-functional-theory (DFT) calculations, we further propose realizing 2D topological chiral p-wave superconductors experimentally by depositing transition metal adatoms on superconducting Pb. The type of magnetic ion can be varied to access different strengths of the magnetic moment. In the case of Fe adatoms on a Pb (111) surface, we show that strong magnetic order in general leads to an odd number of 2D Fermi surface segments. As a consequence the proximity-induced superconducting phases can have nonzero Chern numbers and chiral Majorana edge modes. Below we first present our model system results which bring out a number of generic features of superconductor surfaces with ferromagnetically ordered magnetic adatoms, and then we summarize our DFT results by focusing on the specific case of
Fe adatoms on a Pb (111) surface.

To render the analytical calculations tractable, we consider a Hamiltonian that models only the surface layer of a bulk 3D $s$-wave superconductor on which the $s$-wave superconducting order parameter $\Delta$ is induced from the bulk. On this superconducting layer, we model the magnetic impurities as classical spins whose only interaction with the electrons in the superconductor is through Zeeman-like couplings.~\cite{Yu65,Shiba68,Rusinov68} Employing a tight-binding description on a 2D square lattice $\Lambda$ that is spanned by the primitive lattice vectors $\hat{\bs{e}}_1$ and $\hat{\bs{e}}_2$, we consider the 
mean-field Hamiltonian
\begin{equation}
\begin{split}
H=&\sum_{\bs{r}\in\Lambda}
\left[
t \left(c^\dagger_{\bs{r}}c_{\bs{r}+\hat{\bs{e}}_1}+c^\dagger_{\bs{r}}c_{\bs{r}+\hat{\bs{e}}_2}\right)
-\frac{\mu}{2} c^\dagger_{\bs{r}}c_{\bs{r}}+\Delta\, c^\dagger_{\bs{r},\uparrow}c^\dagger_{\bs{r},\downarrow}
\right.
\\
&\qquad\left.
+\mathrm{i}\,\alpha\left(c^\dagger_{\bs{r}}\sigma_2c_{\bs{r}+\hat{\bs{e}}_1}
-c^\dagger_{\bs{r}}\sigma_1c_{\bs{r}+\hat{\bs{e}}_2}\right)
+\mathrm{h.c.}
\right]
\\
&+J\sum_{\bs{r}\in\Lambda^*}c^\dagger_{\bs{r}}\sigma_3 c_{\bs{r}}.
\label{eq: Hamiltonian}
\end{split}
\end{equation}
Here, $c^\dagger_{\bs{r}}=(c^\dagger_{\bs{r},\uparrow},c^\dagger_{\bs{r},\downarrow})$
is a spinor of the creation operators for electrons at site $\bs{r}$ with spin $\uparrow, \downarrow$,
and $\sigma_{i},\ i=1,2,3,$ are the three Pauli matrices.
We denote by $t$ the nearest neighbor hopping integral in the superconductor, $\mu$ the chemical potential, and $\alpha$ the strength of the  Rashba spin-orbit coupling. 
Classical magnetic moments ferromagnetically aligned normal to the plane in the $\sigma_3$ direction are positioned on a sublattice $\Lambda^*$ of $\Lambda$ and are exchange-coupled to electrons via the term proportional to $J$. Drawing experience from the 1D situation and the ab-initio calculations presented below, the physically relevant hierarchy of energy scales that we consider here is given by
 \begin{equation}
 t\gg J\gg \alpha\sim\Delta.
 \label{eq: limit}
 \end{equation} 
 
\emph{Dense impurity limit} ---
As a warmup, it is instructive to consider the simplest situation in which each lattice site is coupled to a magnetic moment, i.e., $\Lambda=\Lambda^*$.  
This limit is representative of system with 
self-assembled islands of magnetic adatoms.
Its consideration allows us to highlight the difference
 between the regime Eq.~\eqref{eq: limit}, and that of small $J$ that has been studied previously.~\cite{Sato09,Sau10, Alicea10}
 In particular, it is possible to access a phase with Chern number 2 in the large $J$ limit. 
For the case $\Lambda=\Lambda^*$ and $\alpha=0$, the Hamiltonian~\eqref{eq: Hamiltonian}
has gapless lines in momentum space defined by 
\begin{equation}
\varepsilon_{\bs{k}}=\pm\sqrt{J^2-\Delta^2},
\label{eq: gapless condition 1x1}
\end{equation}
 where $\varepsilon_{\bs{k}}=2t(\cos\,k_1+\cos\,k_2)-\mu$. Adding spin-orbit coupling $\alpha\neq0$ will generically lift these degeneracy lines and gap the spectrum except at the four inversion-symmetric momenta [$\bs{k}_{ij}=(1-i,1-j)\pi/2$ with $i,j=\pm$] at which the spin-orbit coupling vanishes and at which phase transitions between different Chern number phases occur. At each of these momenta, the condition~\eqref{eq: gapless condition 1x1} is met for two values of the chemical potential
\begin{equation}
\mu_{i,j,\lambda}=-2(i+j)t+\lambda\sqrt{J^2-\Delta^2},\qquad i,j,\lambda=\pm.
\label{eq: condition for gapless point}
\end{equation}
Around each of these gapless points in the three-dimensional $\mu$-$\bs{k}$ parameter space, we can reduce the Hamiltonian to an effective two-band model and expand it to linear order in the deviations $\delta\bs{k}$ from $\bs{k}_{ij}$ and the deviations $\delta\mu$ from $\mu_{i,j,\lambda}$ yielding
\begin{equation}
H^{\mathrm{eff}}_{i,j,\lambda}
=-\frac{\Delta\alpha}{J}\left(j\,\delta k_2\sigma_1-i\,\delta k_1\sigma_2\right)+\lambda\sqrt{1-\frac{\Delta^2}{J^2}}\,\delta\mu\,\sigma_3.
\label{eq: effective 2-band Hamiltonian}
\end{equation} 
Here, the Pauli matrices act on a subspace defined by the two bands that satisfy condition~\eqref{eq: condition for gapless point}.
In the $\mu$-$\bs{k}$ space, Hamiltonian~\eqref{eq: effective 2-band Hamiltonian}
represents a source of Berry flux that corresponds to a unit topological charge $i\times j\times \lambda=\pm1$.
As the chemical potential ramps through the gapless point at $\mu_{i,j,\lambda}$, the Chern number changes by $i\times j\times \lambda$.
Let us assume $2t>\sqrt{J^2-\Delta^2}$, so that the critical chemical potentials are in the order
$\mu_{-,-,-}<\mu_{-,-,+}<\mu_{-,+,-}=\mu_{+,-,-}<\mu_{-,+,+}=\mu_{+,-,+}<\mu_{+,+,-}<\mu_{+,+,+}$.
Trivially, for $\mu<\mu_{-,-,-}$ the Chern number is zero. By increasing $\mu$, the system thus exhibits phases with the Chern numbers
\begin{equation}
C=\ -1,\ 0,\ +2,\ 0,\ -1,
\end{equation}
where the phase transitions occur at the respective $\mu_{i,j,\lambda}$.
Hence, with homogeneous magnetization, the superconductor may already exhibit Chern number equal to 2. In cases where magnetic impurities are spaced more sparsely, i.e., if $\Lambda^*$ is a sublattice of $\Lambda$, even higher Chern numbers can be obtained.

\begin{figure}
\centering
\includegraphics[width=\linewidth]{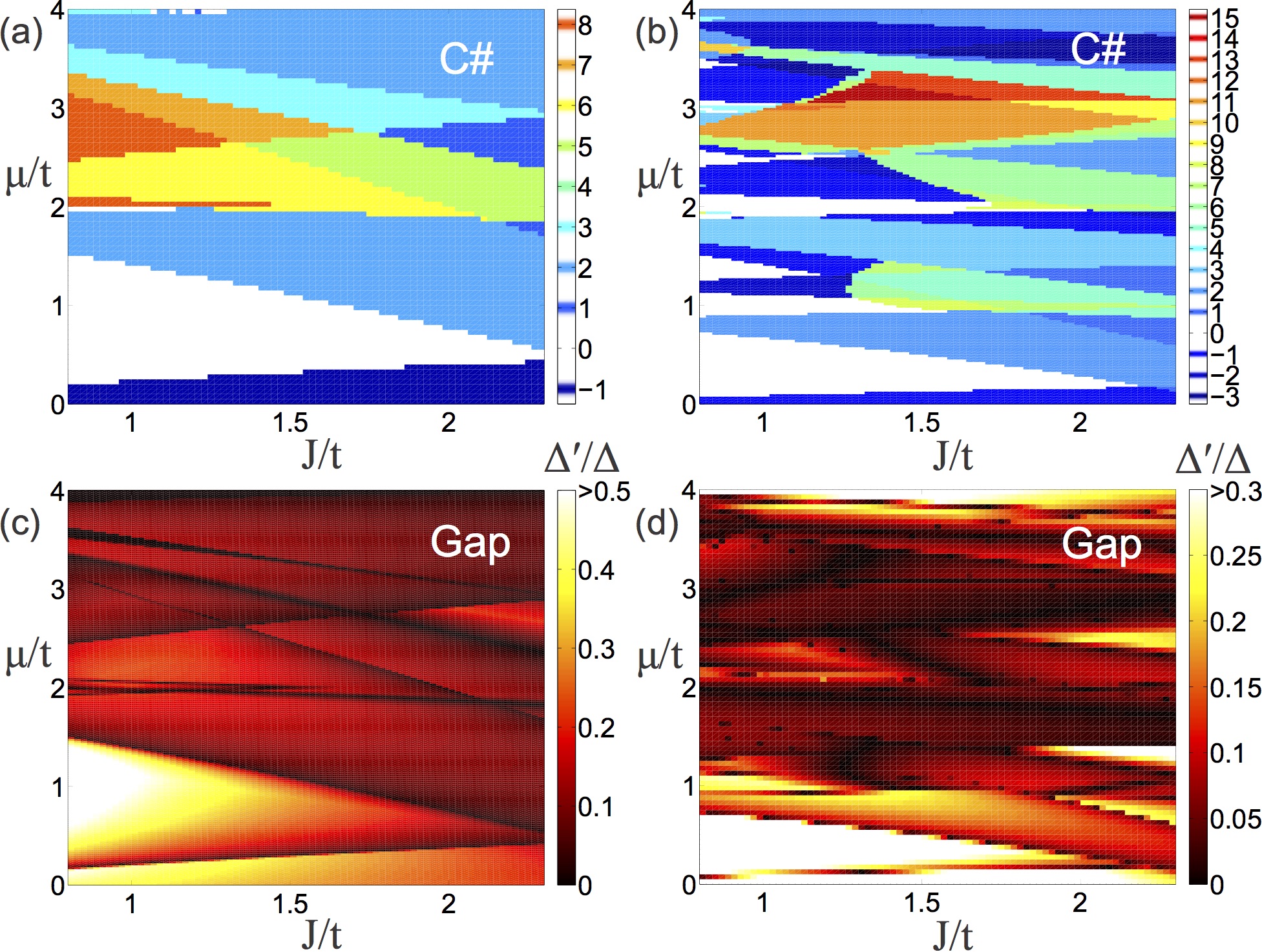}
\caption{
Phase diagrams for Hamiltonian~\eqref{eq: Hamiltonian}, in terms of Chern numbers [(a) and (b)] and superconducting gaps [(c) and (d)] as a function of the chemical potential $\mu$ and the strength of the magnetic moments $J$. The panels (a) and (c) [(b) and (d)] correspond to the case of one magnetic adatom every 2$\times$2 (3$\times$3) lattice sites. In these plots, $\Delta=0.06t$ and $\alpha=0.1t$; $\mu$ has been shifted such that $\mu=0$ implies the chemical potential lying in the center between the two spin-split band bottoms.}
\label{fig: lattice phase diagrams}
\end{figure}

We solved Hamiltonian~\eqref{eq: Hamiltonian} numerically for the case of one magnetic impurity every $2\times2$ and $3\times 3$ plaquettes of the square lattice, and show the phase diagrams in Fig.~\ref{fig: lattice phase diagrams}.
From the phase diagrams, we can read off three general features:
i) at small chemical potentials around the band bottom, where the Fermi wavelengths are larger than or comparable to the lattice spacing of $\Lambda^*$, the sequence of Chern numbers $(-1,0,+2)$ always occurs when $J/t$ is not too large -- this universal feature corresponds to the dense limit that we have discussed above;
ii) at larger chemical potentials, more than two Fermi surfaces can exist in the reduced Brillouin zone defined by $\Lambda^*$ as the Fermi wavelengths are significantly smaller than the lattice spacing of $\Lambda^*$, as a consequence higher Chern numbers can occur [e.g. 8 in Fig.~\ref{fig: lattice phase diagrams}(a) and 15 in Fig.~\ref{fig: lattice phase diagrams}(b)] but the trade-off is an overall smaller induced gap;
iii) the phases with different Chern numbers are generally separated by lines (in the 2D parameter space) defined by conditions similar to Eq.~\eqref{eq: condition for gapless point}, and across each specific line the change of Chern numbers is a constant determined essentially the same way as in our preceding analysis.

\emph{Dilute impurity limit} --- 
To better understand the dilute impurity limit, we complement our results on the lattice by a calculation in which we treat the underlying 2D superconductor in the continuum limit and consider sparsely distributed Shiba impurities that are arranged in a square lattice on top of it. This allows us to derive an effective two-band model for the hybridizing Shiba states. This effective Hamiltonian represents a chiral $p$-wave superconductor in the appropriate parameter regime. 
 
\begin{subequations}
The strategy of our derivation is inspired by the calculation for 1D Shiba chains of Ref.~\cite{Pientka13} (for details, see~\cite{Supp}).  We start from a $4\times 4$ Bogoliubov-de-Gennes (BdG) Hamiltonian
\begin{equation}
H^{\mathrm{BdG}}_{\bs{k}}=
\begin{pmatrix}
\xi_{\bs{k}}&\Delta\\
\Delta^*&-\xi_{\bs{k}}
\end{pmatrix}
\end{equation} 
that acts on $4$-spinor valued wave functions 
$\Psi(\bs{r})=(\psi_\uparrow,\psi_\downarrow,\psi^\dagger_\downarrow,-\psi^\dagger_\uparrow)^{\mathsf{T}}(\bs{r})$, 
$\bs{r}\in\mathbb{R}^2$, with $\xi_{\bs{k}}=\bs{k}^2/(2m)-\mu+\alpha(k_1\sigma_2-k_2\sigma_1)$. In addition, the magnetic impurities are represented by the Hamiltonian
\begin{equation}
H^{\mathrm{imp}}=
-J\,\sum_{\bs{r}^*\in\Lambda^*}
\delta(\bs{r}-\bs{r}^*)
S_3,
\end{equation}
\end{subequations}
where $S_3=\sigma_3\otimes\tau_0$, with $\tau_3$ the identity matrix acting on particle-hole space. If we restrict the wave-function $\Psi(\bs{r})$ to the locations $\bs{r}^*\in\Lambda^*$ of the impurities, we obtain the self-consistency equation
\begin{equation}
\left(E-\mathcal{H}^{\mathrm{BdG}}_{\bs{q}}\right)^{-1}
\,(-J S_3)
\widetilde{\Psi}(\bs{q})
=\widetilde{\Psi}(\bs{q})
\label{eq: consistency Eq dilute}
\end{equation} 
for the Fourier transforms $\widetilde{\Psi}(\bs{q})=\sum_{\bs{r}^*\in\Lambda^*}e^{-\mathrm{i}\bs{q}\cdot\bs{r}^*}\Psi(\bs{r}^*)$, where the momentum $\bs{q}\in[0,2\pi)^2$ now belongs to the  $\Lambda^*$ Brillouin zone and we have set the impurity spacing to unity. 

\begin{subequations}
We need two more steps to reduce Eq.~\eqref{eq: consistency Eq dilute} to an effective two-band model for the Shiba states, assuming they are deep in the superconducting gap and dilute compared with the Fermi wavelength.
First, the left-hand side is expanded to linear order in the energy $E$ to cast the equation in the form of the time-independent Schroedinger equation. Second, we project the effective Hamiltonian into the eigenstates of an isolated Shiba impurity on every site, given by $\Psi_+=(1,0,1,0)^{\mathsf{T}}/\sqrt{2}$ and $\Psi_-=(0,1,0,-1)^{\mathsf{T}}/\sqrt{2}$~\cite{Pientka13}. We obtain the effective two-band Hamiltonian
\begin{equation}
H^{\mathrm{eff}}_{\bs{q}}
=\frac{1}{\eta+d_{0,\bs{q}}}
\left[
\left(1-\eta-d_{0,\bs{q}}\right)\sigma_3
-d_{2,\bs{q}}\sigma_1+d_{1,\bs{q}}\sigma_2
\right],
\end{equation}
where
\begin{eqnarray}
d_{0,\bs{q}}&=&
\frac{J}{2}\sum_{\bs{r}^*\neq 0}e^{-\mathrm{i}\bs{q}\cdot\bs{r}^*} f_+(|\bs{r}^*|),
\\
d_{i,\bs{q}}&=&\mathrm{i}\,
\frac{J}{2}
\sum_{\bs{r}^*\neq 0}
e^{-\mathrm{i}\bs{q}\cdot\bs{r}^*} \frac{r^*_i}{|\bs{r}^*|}f_-(|\bs{r}^*|),
\end{eqnarray}
are defined in terms of the functions
\begin{equation}
f_\pm(r)
=
\frac{e^{-r/\xi}}{v_{\mathrm{F}}\sqrt{2\pi r}}
\sum_{\lambda=\pm} (\pm1)^{(1-\lambda)/2}
\sqrt{k_{\mathrm{F}}^{(\lambda)}}\cos\left(k_{\mathrm{F}}^{(\lambda)}\, r-\frac{\pi}{4}\right),
\end{equation}
\label{eq: effective 2-band H}
\end{subequations}
In the equation above, $\xi$ is the coherence length of the 2D superconductor (without the magnetic impurities), $v_{\mathrm{F}}=\sqrt{\alpha^2+2\mu/m}$ is its Fermi velocity, $k_{\mathrm{F}}^{(\pm)}=\sqrt{2m\mu+\alpha^2m^2}\mp m\alpha$ are the Fermi wave-vectors for its two spin-split bands, and $\eta=J [k_{\mathrm{F}}^{(+)}+k_{\mathrm{F}}^{(-)}]/4v_F$ is a dimensionless parameter. In the ``deep Shiba limit'' in which the projection in the states $\Psi_+$ and $\Psi_-$ is justified, we have $\eta\sim1$.

\begin{figure}
\centering
\includegraphics[width=\linewidth]{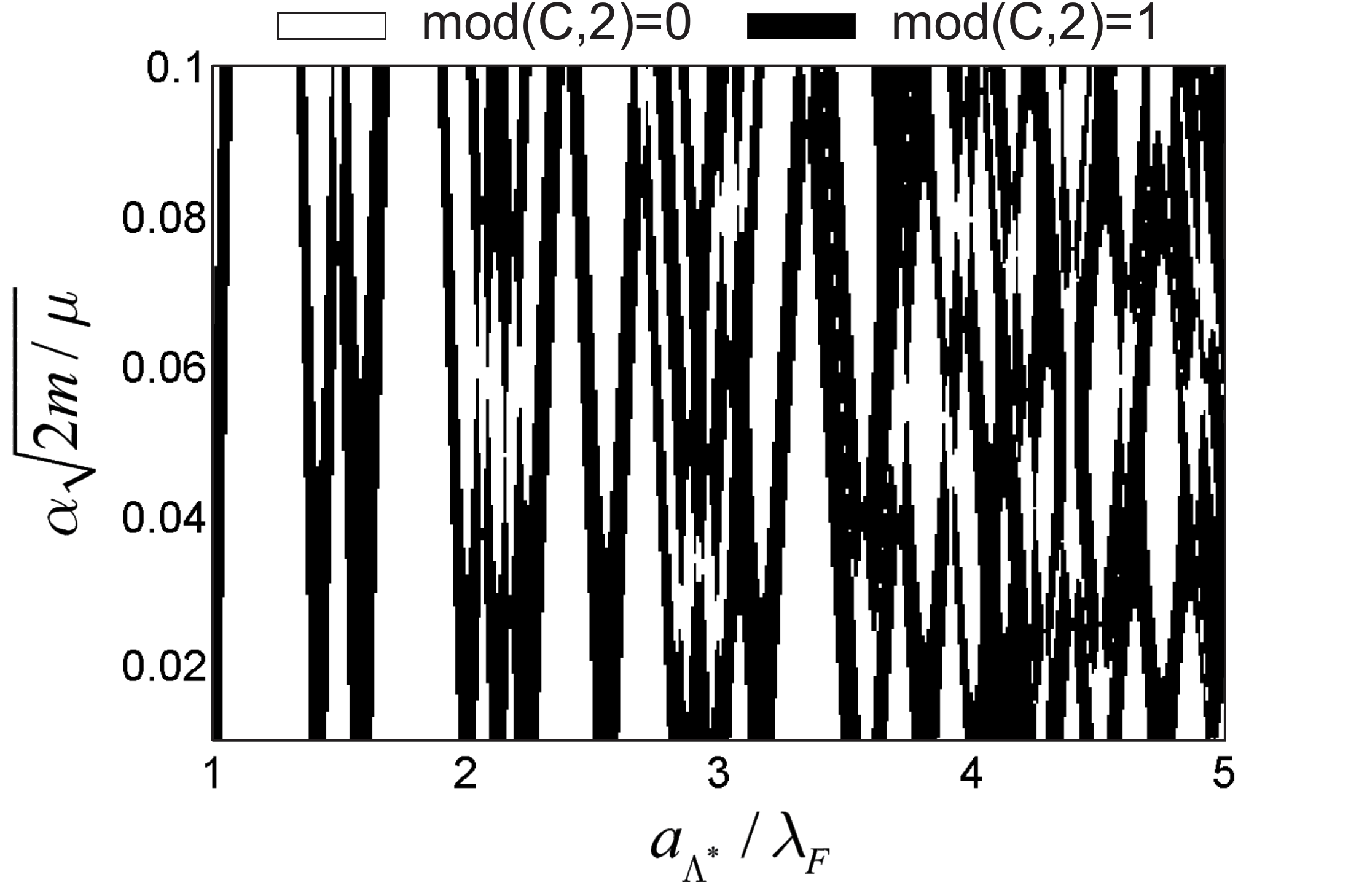}
\caption{
Parity of the Chern number for the BdG band structure 
of an $s$-wave superconductor decorated with dilute Shiba impurities, following Eq.~\eqref{eq: Chern number parity}. 
Here, $a_{\Lambda^*}$ is the lattice spacing of the Shiba impurities, $\lambda_{\mathrm{F}}$ is the Fermi wavelength in the limit $\alpha\to 0$ and 
the Hamiltonian~\eqref{eq: effective 2-band H} was evaluated with $\xi=30\lambda_F$.
}
\label{fig: chern number parity}
\end{figure}

The Hamiltonian~\eqref{eq: effective 2-band H} represents the effective superconductor formed by the Shiba bound states within the gap of the underlying $s$-wave superconductor. 
Similar to the case of the effective two-band model in Eq.~\eqref{eq: effective 2-band Hamiltonian}, this Hamiltonian can have nodal points in $\bs{k}$-$\mu$-space at which the Chern number changes. However, unlike in Eq.~\eqref{eq: effective 2-band Hamiltonian}, the nodes can occur at any point in the Brillouin zone, making an analytic treatment intractable. In addition, the validity of Hamiltonian~\eqref{eq: effective 2-band H} requires a self-consistency that permits the low-energy expansion of Eq.~\eqref{eq: consistency Eq dilute}. Therefore, instead of computing the Chern numbers in an extended parameter space, we focus on information that can be obtained at special points of the Brillouin zone at infinitesimal energy. To that end, observe that Hamiltonian~\eqref{eq: effective 2-band H} has $C_4$ rotational symmetry. Thus, any gap closing at points other than the $C_4$-symmetric momenta $\bs{k}=(0,0)$ and $\bs{k}=(\pi,\pi)$ changes the Chern number by an even integer due to the symmetry-imposed multiplicity of the nodal points. By expanding the Hamiltonian into Dirac form around the $C_4$-symmetric momenta, we obtain the expression~\cite{Supp}
\begin{equation}
(-1)^C=\mathrm{sgn}
\left[d_{0,(0,0)}\right]\mathrm{sgn}\left[d_{0,(\pi,\pi)}\right]
\quad (\eta=1)
\label{eq: Chern number parity}
\end{equation}
for the parity of the Chern number. The numerical evaluation of this equation is shown in Fig.~\eqref{fig: chern number parity} in the form of a phase diagram.

We have also performed calculations for magnetic orders other than simple ferromagnetism.
In particular, the case where the magnetic configurations corresponds to 2D helices is related to previous studies on 1D helices.~\cite{Choy11,Nadj-Perge13,Nakosai13,Nakosai13,Klinovaja13,Braunecker13,Vazifeh13,Po13,Pientka13b,Kim14,Li14-a} We obtained criteria for such a system to be fully gapped by proximity effect, and found that the fully-gapped superconducting phases can be generically topologically nontrivial. The results and phase diagrams are presented in the supplementary material.~\cite{Supp}

\begin{figure}
\centering
\includegraphics[width=\linewidth]{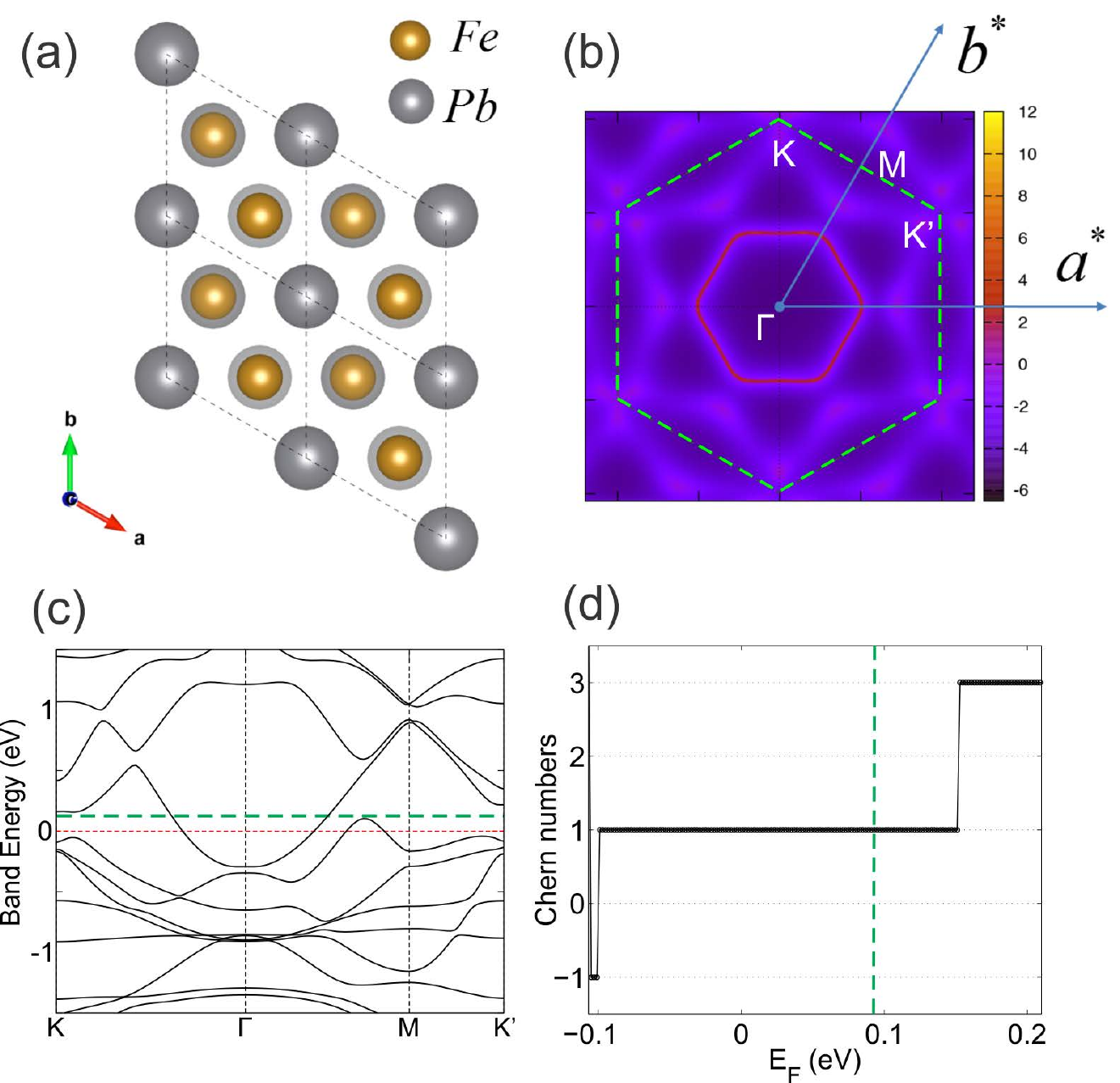}
\caption{
Electronic structure from DFT calculations for Fe adatoms on Pb (111) surface.
(a) The atomic configuration: Fe adatoms form a buckled a honeycomb lattice on the Pb (111) surface.
(b) Fermi surface structure of the Fe $d$-orbitals for the Fermi energy indicated by the dashed lines in panel (c) and (d) (see~\cite{Supp} for cases with different Fermi energies). A single Fermi surface is found around the $\Gamma$ point. 
(c) DFT band structure of the Fe $d$-orbitals with spin-orbit coupling.
(d) Chern number of the BdG bands when a small $s$-wave superconductivity ($\Delta=0.01eV$ in this calculation) is induced in the system.
}
\label{fig: dft results}
\end{figure}

In closing, we complement our simple model considerations with a specific material proposal to realize a superconductor with nonzero Chern numbers. For that, we consider transition metal atoms, in particular Fe, deposited on the (111) surface of Pb, a strong type-II superconductor. The same combination of materials, but a different surface of Pb, was used in the experimental realization for the 1D $p$-wave superconductor.~\cite{Nadj-Perge14} The Pb atoms on the (111) surface form a triangular lattice. Through ab-initio calculations that assume a ferromagnetic alignment of the Fe magnetic moments and include spin-orbit coupling, we compared the relaxation energies for various densities and arrangements of Fe adatoms on the Pb (111) surface, and found that a deposition with one Fe atom in each triangular plaquette is particularly favorable (for details, see~\cite{Supp}). In this case, the Fe atoms form a honeycomb lattice in which the atoms sit at different heights in each sublattice [see Fig.~\ref{fig: dft results}(a)]. We further performed DFT calculations of the electronic structure. The resulting (not spin-degenerate) Fermi surface and band structure restricted to the Fe $d$-orbitals is shown in Fig.~\ref{fig: dft results}(b) and (c). Critically, we find an odd number of Fermi surfaces. The Chern number of the corresponding BdG Hamiltonian is indeed nonzero over a large range of the chemical potential [see Fig.~\ref{fig: dft results}(d)].

In conclusion, we have proposed a versatile platform for realizing chiral superconductors in 2D. We have obtained analytically the topological phase diagram (Chern number and gaps of the superconductor) of the dilute and dense limit, and numerically evaluated the phase diagram in the intermediate regime. We then showed through a more realistic ab-initio calculation that ferromagnetically ordered Fe atoms on the (111) surface of Pb in the dense limit  could give rise to a chiral superconductor. 
The presence of a 2D chiral superconductor could be established experimentally by tunneling into the chiral Majorana modes, whose number is equal to the Chern number of the phase, and which would take place only on the edge of a 2D thin island of Fe on the surface of Pb. 

\emph{Note} During the completion of this work, a similar proposal for realizing chiral superconductivity appeared on the arxiv.~\cite{Roentynen14} Where the present manuscript overlaps with~\cite{Roentynen14}, in the dilute limit, our calculation agrees with the one presented in~\cite{Roentynen14}. Our paper also further presents the non-dilute limit and the ab-initio results for realistic materials. 

\begin{acknowledgments}
We thank F. Pientka for discussions. 
This work was supported by DARPA SPAWARSYSCEN Pacific N66001-11-1-4110, and ONR-N00014-14-1-0330. BAB acknowledges support from NSF CAREER DMR-0952428, ONR-N00014-11-1-0635, MURI-130-6082, DARPA under SPAWAR Grant No.: N66001-11-1-4110, the Packard Foundation, and a Keck grant.  JL acknowledges support from Swiss National Science Foundation. AHM acknowledges support from the Welch Foundation under grant TBF1473. AY acknowledges support from NSF–MRESEC-DMR-0819860 and NSF-DMR-1104612.
\end{acknowledgments}

\clearpage
\newpage

\appendix
\begin{widetext}
\begin{center}
\large
\textbf{
Supplemental Material for 
``A novel platform for two-dimensional chiral topological superconductivity''}
\normalsize
\end{center}

\tableofcontents

\section{Derivation of the effective Hamiltonian in the dilute Shiba limit}

\subsection{General strategy}
\label{sec:genform}

We want a general formalism to find the low energy excitations and effective Hamiltonian of the following Hamiltonian
\begin{align}
  H = {H}_0 + H_1,\quad H_1=\sum_{n\in c} V_n \delta(r-r_n),
\end{align}
where $H_0$ is the original Hamiltonian which is gapped ($\approx\Delta$) around zero energy, $V_n$ are matrices associated with the inner degrees of freedom (spin, particle-hole ect.) which can induce in-gap states, and $r$ is implicitly $d$-dimensional (also $n$, $m$, $k$, $q$, $R$, $n_b$ below).

We start with the Schr\"odinger equation
\begin{align}
  (H_0 + H_1)\Psi = E\Psi, \quad (E<\Delta).
\end{align}
It follows that
\begin{align}\label{eq:G0H1}
  G_0(E) H_1 \Psi = \Psi,
\end{align}
where $G_0(E)=(E-H_0)^{-1}$ is the Green function.

Because $H_1$ is composed of delta-functions for a small set $c$, $G_0 H_1$ is nonzero only in the columns corresponding to $c$, thus Eq.~\ref{eq:G0H1} is equivalent to
\begin{align}\label{eq:G0Vi}
  \sum_{n\in c} G_0(E;r,r_n) V_n \psi(r_n) = \psi(r).
\end{align}
In the simplest case, if $c$ contains only one single point, labeled by $0$, then Eq.~\ref{eq:G0Vi} implies
\begin{align}
  G_0(E;r_0,r_0) V_0 \psi(r_0) = \psi(r_0).
\end{align}
The energy of the excitations can be obtained by solving
\begin{align}
  &\text{Det}[\mathds{1}-G_0(E;r_0,r_0) V_0] = 0, \\
  &G_0(E;r_0,r_0) = G_0(E;0) = \int dk \; G_0(E;k).
\end{align}
In more complicated cases, the following equation, again implied by Eq.~\ref{eq:G0Vi}, can serve as the starting point to extract an effective Hamiltonian
\begin{align}
  &\forall m\in c: \quad \sum_{n\in c} G_0(E;r_m,r_n) V_n \psi(r_n) = \psi(r_m), \label{eq:G0Vi2}\\
  &G_0(E;r_m,r_n) = G_0(E;r_m-r_n) = \int dk \; G_0(E;k) e^{ik(r_m-r_n)}.
\end{align}

In addition if $\forall n:\; V_n = V_0$ and $r_n = n R$ ($n\in \mathds{Z}^d$ with $d$ the dimension), Eq.~\eqref{eq:G0Vi2} can be transformed to $k$-space:
\begin{align}
  \psi(q) &= \sum_{m} e^{-iqr_m} \psi(r_m) \qquad (q\in [-\pi/R,\pi/R)) \\
  &= \sum_{m,n} e^{-iqr_m} G_0(E;r_m-r_n) V_0 \psi(r_n) \\
  &= \sum_{m,n} \int dk \; G_0(E;k) e^{i(k-q)(r_m-r_n)} V_0 e^{-iqr_n} \psi(r_n) \\
  &= \int dk \; G_0(E;k) \left[\sum_{m'} e^{i(k-q)m'R}\right] V_0 \left[\sum_{n} e^{-iqr_n} \psi(r_n)\right] \\
  &= \int dk \; G_0(E;k) \left[\sum_{n_b} \delta(k-q-2n_b\pi/R)\right] V_0 \psi(q) \\
  &= G_0(E;q) V_0 \psi(q),
\end{align}
where
\begin{align}
  G_0(E;q) := \sum_{n_b}  G_0(E;q+2n_b\pi/R).
\end{align}
$n_b$ can be interpreted as the band index when the $k$-space is folded into the Brillouin zone defined by $[-\pi/R,\pi/R)$.

\subsection{Shiba lattice in the 2D Rashba superconductor}

We start with the BdG Hamiltonian
\begin{align}
  &H_{BdG}(k) =
  \begin{pmatrix}
    \xi_k & \Delta_k \\
    \Delta_k^\dagger & -\xi_k
  \end{pmatrix}, \\
  &\xi_k = \frac{\hbar^2}{2m^*}k^2 - \mu + \alpha (k_x \sigma_y - k_y \sigma_x)\,, \\
  &\Delta_k  = \Delta_s\sigma_0 + \delta_p(k_x \sigma_y - k_y \sigma_x).
\end{align}
First we make the Hamiltonian dimensionless by defining
\begin{align}
  k_{\mu} = \sqrt{2m^*\mu/\hbar^2}, \quad
  \tilde{\bm{k}} = \bm{k}/k_{\mu}, \\
  l_{\alpha} = \hbar^2/(2m^*\alpha), \quad
  \tilde{\alpha} = 1/(l_{\alpha}k_{\mu}), \\
  l_p = \hbar^2/(2m^*\delta_p), \quad
  \Delta_p = 1/(l_p k_{\mu}), \\
  \tilde{H} = H/\mu, \quad
  \tilde{\Delta}_s = \Delta_s/\mu,
\end{align}
and then omit the tildes, the bulk Hamiltonian becomes
\begin{align}
  H_{BdG}(\bm{k}) = \;&\tau_z\otimes[(k^2-1)\sigma_0+\alpha(k_x \sigma_y - k_y \sigma_x)] \nonumber\\
  + &\tau_x\otimes[\Delta_s\sigma_0 + \Delta_p(k_x \sigma_y - k_y \sigma_x)].
\end{align}
In this dimensionless Hamiltonian, energies are measured in units of $\mu$ (assumed to be large compared with other energy scales), lengths (wave vectors) are measured in units of the Fermi wavelength (wave vector).

Next we go to the basis where the normal state Hamiltonian is diagonalized 
\begin{align}
  \tilde{H}_{BdG}(k) &= U(\varphi_k)^\dagger H_{BdG}(\bm{k}) U(\varphi_k) \\
  &= \tau_z\otimes[(k^2-1)\sigma_0+\alpha k \sigma_z] + \tau_x\otimes[\Delta_s\sigma_0 + \Delta_p k \sigma_z] \\
  &= [E_+(k)\tau_z+\Delta_+(k)\tau_x] \oplus [E_-(k)\tau_z+\Delta_-(k)\tau_x],
\end{align}
where
\begin{align}
  &U(\varphi_k) = \tau_0\otimes\frac{1}{\sqrt{2}}
  \begin{pmatrix}
    1 & -1 \\
    ie^{i\varphi_k} & ie^{i\varphi_k}
  \end{pmatrix}, \\
  &E_\pm(k) = k^2-1\pm \alpha k, \\
  &\Delta_\pm(k) = \Delta_s \pm \Delta_p k.
\end{align}

According to the general formalism (Sec.~\ref{sec:genform}), we need to evaluate
\begin{align}
  G_0(E,\bm{r}) &= \int d\bm{k} \; G_0(E,\bm{k}) e^{i\bm{k}\cdot\bm{r}} \\
  &= \int_0^{\infty} \frac{k\,dk}{2\pi} \int_0^{2\pi} \frac{d\varphi_k}{2\pi} \; U(\varphi_k) [E-\tilde{H}_{BdG}(k)]^{-1} U(\varphi_k)^\dagger e^{i k r \cos(\varphi_k-\varphi_r)}\\
  &= R(\varphi_r) \left[ \int_0^{2\pi} \frac{d\varphi'_k}{2\pi} \; U(\varphi'_k) \Gamma(E,r\cos\varphi'_k) U(\varphi'_k)^\dagger \right] R(\varphi_r)^\dagger, \label{eq:G0int}
\end{align}
where
\begin{align}
  &\Gamma(E,x) = \int_0^{\infty} \frac{k\,dk}{2\pi} e^{i k x} [E-\tilde{H}_{BdG}(k)]^{-1}, \\
  &R(\varphi_r) = \tau_0\otimes
  \begin{pmatrix}
    1 &0 \\
    0 & e^{i\varphi_r}
  \end{pmatrix}.
\end{align}

To evaluate $\Gamma(E,x)$, we first perform the integral
\begin{align}
  \Gamma_{b=\pm}(E,x) &= \int_0^{\infty} \frac{k\,dk}{2\pi} e^{i k x} [E-E_b(k)\tau_z-\Delta_b(k)\tau_x]^{-1} \\
  &\simeq \rho_b \int dE_b \, e^{i [k_F^{(b)}+E_b/v_F] x} \,\frac{E+E_b\tau_z+\Delta_b\tau_x}{E^2-E_b^2-\Delta_b^2},
\end{align}
where $v_F = 2k_0$ ($k_0\equiv\sqrt{1+\alpha^2/4}$) is the Fermi velocity (the same for $\pm$ bands), $k_F^{(\pm)} = k_0\mp\alpha/2$ is the Fermi wave vector for $+/-$-band, $\rho_b = k_F^{(b)}/2\pi v_F$ is the density of states for $b$-band at the Fermi energy, and $\Delta_b=\Delta_b(k_F^{(b)})$ is the pairing gap at the Fermi energy. Following Pientka et al. \cite{pientka_2013}, we have
\begin{align}
 &\int dE_b \, e^{i E_b x/v_F} \,\frac{1}{E_b^2+a^2} = \frac{\pi}{a}e^{-a|x|/v_F}, \\
 &\int dE_b \, e^{i E_b x/v_F} \,\frac{E_b}{E_b^2+a^2} \frac{\omega_D^2}{E_b^2+\omega_D^2} \nonumber\\
 &= \frac{i\pi \omega_D^2}{\omega_D^2-a^2}\,\text{sgn}(x)(e^{-a|x|/v_F}-e^{-\omega_D|x|/v_F}),
\end{align}
where $a=\sqrt{\Delta_b^2-E^2}$ is a short-hand notation, and $\omega_D$ is the Debye frequency which will be sent to $+\infty$ in the end. Then
\begin{align}
  &\Gamma_{b} = \gamma_{0b}+ \gamma_{1b}\tau_x+ \gamma_{3b}\tau_z, \\
  &\Gamma = \Gamma_{+} \oplus \Gamma_{-} =
  \begin{pmatrix}
     \gamma_{0+}+ \gamma_{3+} & 0 &  \gamma_{1+} & 0 \\
    0 &  \gamma_{0-}+ \gamma_{3-} & 0 &  \gamma_{1-} \\
     \gamma_{1+} & 0 &  \gamma_{0+}- \gamma_{3+} & 0 \\
    0 &  \gamma_{1-} & 0 &  \gamma_{0-}- \gamma_{3-}
  \end{pmatrix},
\end{align}
where
\begin{align}
  & \gamma_{0b} = - \pi\rho_b \, \frac{E}{\sqrt{\Delta_b^2-E^2}} \, e^{i k_F^{(b)}x-\sqrt{\Delta_b^2-E^2}|x|/v_F}, \\
  & \gamma_{1b} = - \pi\rho_b \, \frac{\Delta_b}{\sqrt{\Delta_b^2-E^2}} \, e^{i k_F^{(b)}x-\sqrt{\Delta_b^2-E^2}|x|/v_F}, \\
  & \gamma_{3b} = - \pi\rho_b \, \frac{i\omega_D^2}{\omega_D^2+E^2-\Delta_b^2}\,\text{sgn}(x) e^{i k_F^{(b)}x} (e^{-\sqrt{\Delta_b^2-E^2}|x|/v_F}-e^{-\omega_D|x|/v_F}).
\end{align}
Next we find
\begin{align}
  U(\varphi_k) \,\Gamma\, U(\varphi_k)^\dagger =
  \frac{1}{2}\,
  \begin{pmatrix}
    A_{+} & -ie^{-i\varphi_k}A_{-} & C_+ &  -ie^{-i\varphi_k}C_{-} \\
    ie^{i\varphi_k}A_{-} & A_+ &  ie^{i\varphi_k}C_{-} & C_+ \\
    C_+ &  -ie^{-i\varphi_k}C_{-} & B_{+} & -ie^{-i\varphi_k}B_{-} \\
    ie^{i\varphi_k}C_{-} & C_+ &  ie^{i\varphi_k}B_{-} & B_+
  \end{pmatrix},
\end{align}
where
\begin{align}
  &A_{\pm} = (\gamma_{0+} + \gamma_{3+}) \pm (\gamma_{0-} + \gamma_{3-}), \\
  &B_{\pm} = (\gamma_{0+} - \gamma_{3+}) \pm (\gamma_{0-} - \gamma_{3-}), \\
  &C_{\pm} = \gamma_{1+} \pm \gamma_{1-}.
\end{align}

The integral Eq.~\eqref{eq:G0int} breaks down to the following basic one
\begin{align}
  I_l(z) &=i^l \int_{-\pi/2}^{\pi/2} d\varphi \; e^{i (l\varphi + z \cos\varphi)} \quad (l=0,\pm1) \\
  &=i^l \int_{-\pi/2}^{\pi/2} d\varphi \; e^{i l\varphi} \left[\sum_{n=-\infty}^{+\infty} i^n J_n(z) e^{in\varphi}\right] \quad (\text{Jacobi-Anger expansion}) \\
  &=\sum_{n=-\infty}^{+\infty} i^{l+n} J_n(z) \int_{-\pi/2}^{\pi/2} d\varphi \; e^{i (l+n)\varphi} \\
  &=\pi J_{-l}(z) + 2i\sum_{m=-\infty}^{+\infty} \frac{J_{2m+1-l}(z)}{2m+1},
\end{align}
where $J_n(z)$ is the the $n$-th order Bessel function of the first kind. Expressed in terms of $I_l(z)$, we have
\begin{align}
  &\int_0^{2\pi} \frac{d\varphi_k}{2\pi} \, i^l e^{i l\varphi_k} \gamma_{0b}(\cos\varphi_k)
  = - \frac{\rho_b}{2} \, \frac{E}{\sqrt{\Delta_b^2-E^2}}
  \left[ I_l(z_{b}) + (-1)^l I_l(-z_{b}^*) \right], \\
  &\int_0^{2\pi} \frac{d\varphi_k}{2\pi} \, i^l e^{i l\varphi_k} \gamma_{1b}(\cos\varphi_k)
  = - \frac{\rho_b}{2} \, \frac{\Delta_b}{\sqrt{\Delta_b^2-E^2}}
  \left[ I_l(z_{b}) + (-1)^l I_l(-z_{b}^*) \right], \\
  &\int_0^{2\pi} \frac{d\varphi_k}{2\pi} \, i^l e^{i l\varphi_k} \gamma_{3b}(\cos\varphi_k) \\
  &\quad= - \frac{\rho_b}{2} \, \frac{i\omega_D^2}{\omega_D^2+E^2-\Delta_b^2}
  \left[ I_l(z_{b}) - (-1)^l I_l(-z_{b}^*) - I_l(z'_{b}) + (-1)^l I_l(-{z'_{b}}^*) \right],\\
  &z_{b}\equiv k_F^{(b)}r+i\sqrt{\Delta_b^2-E^2}r/v_F,  \quad z'_{b}\equiv k_F^{(b)}r+i\omega_D r/v_F.
\end{align}

First we look at the limit $r\rightarrow 0$, such that $z_b, z'_b \rightarrow 0$. In this case
\begin{align}
  I_0(z=0) = \pi, \quad I_{\pm1}(z=0) = \pm 2i.
\end{align}
We find
\begin{align}
  &G_0(E,\bm{r}\rightarrow 0)
  = R(\varphi_r)
  (\tilde{A}\,\tau_0\otimes\sigma_0 + \tilde{C}\,\tau_x\otimes\sigma_0)
  R(\varphi_r)^\dagger \\
  &\qquad\qquad\qquad= \tilde{A}\,\tau_0\otimes\sigma_0 + \tilde{C}\,\tau_x\otimes\sigma_0, \\
  &\tilde{A} \equiv \frac{1}{2}\int_0^{2\pi} \frac{d\varphi_k}{2\pi} (\gamma_{0+}+\gamma_{0-}) \Bigr|_{r\rightarrow 0} 
  = (-\frac{\pi}{2})\sum_{b} \frac{\rho_b E}{\sqrt{\Delta_b^2-E^2}}, \\
  &\tilde{C} \equiv \frac{1}{2}\int_0^{2\pi} \frac{d\varphi_k}{2\pi} (\gamma_{1+}+\gamma_{1-}) \Bigr|_{r\rightarrow 0}
  = (-\frac{\pi}{2})\sum_{b} \frac{\rho_b \Delta_b}{\sqrt{\Delta_b^2-E^2}}.
\end{align}

For $r>0$, we assume $\text{Re}(z)\gg|\text{Im}(z)|$ and $\text{Re}(z)\gg|n^2-1/4|$ (note that $\text{Re}(z) = \pm k_F r$, and $\text{Im}(z) = \sqrt{\Delta_b^2-E^2}r/v_F$ or $\omega_Dr/v_F$ in our integrals), by using the asymptotic form
\begin{align}
  J_n(z) \approx \sqrt{\frac{2}{\pi z}}\cos(z-n\pi/2-\pi/4),
\end{align}
we obtain
\begin{align}
   I_l(z|\text{Re}(z)>0) &\approx \pi J_{-l}(z) + 2i\sqrt{\frac{2}{\pi z}}\sum_{m=-\infty}^{+\infty} \frac{\cos[z-(2m+1-l)\pi/2-\pi/4]}{2m+1} \\
   &= \pi J_{-l}(z) + 2i\sqrt{\frac{2}{\pi z}}\sin(z+l\pi/2-\pi/4)\sum_{m=-\infty}^{+\infty} \frac{(-1)^m}{2m+1} \\
   &= \sqrt{\frac{2\pi}{z}}e^{i(z-\pi/4+l\pi/2)}= i^{l}\sqrt{\frac{2\pi}{z}}e^{i(z-\pi/4)}\,, \\
   I_l(z|\text{Re}(z)<0) &= [I_{-l}(-z^*)]^* \approx  i^{l}\left[\sqrt{\frac{2\pi}{-z^*}}e^{i(-z^*-\pi/4)}\right]^*.
\end{align}
Therefore (assuming $\omega_D \rightarrow +\infty$ after the integral)
\begin{align}
&G_0(E,\bm{r}) = \frac{1}{2}\,
  R(\varphi_r)
  \begin{pmatrix}
    \tilde{A}_{+} - i\tilde{A}_{-}\sigma_y & \tilde{C}_{+} - i\tilde{C}_{-}\sigma_y \\
    \tilde{C}_{+} - i\tilde{C}_{-}\sigma_y &  \tilde{B}_{+} - i\tilde{B}_{-}\sigma_y
  \end{pmatrix}
  R(\varphi_r)^\dagger,
\end{align}
where
\begin{align}
  &\tilde{A}_{+} \equiv \int_0^{2\pi} \frac{d\varphi_k}{2\pi} A_{+} \\
  &\quad \approx -(\frac{E}{\sqrt{\Delta_+^2-E^2}}c_+ F_+ + \frac{E}{\sqrt{\Delta_-^2-E^2}}c_- F_-) + (s_+ F_+ + s_- F_-), \\
  &\tilde{A}_{-} \equiv \int_0^{2\pi} \frac{d\varphi_k}{2\pi} ie^{i\varphi_k} A_{-} \\
  &\quad \approx (\frac{E}{\sqrt{\Delta_+^2-E^2}}s_+ F_+ - \frac{E}{\sqrt{\Delta_-^2-E^2}}s_- F_-) + (c_+ F_+ - c_- F_-),
\end{align}
\begin{align}
  &\tilde{B}_{+} \equiv \int_0^{2\pi} \frac{d\varphi_k}{2\pi} B_{+} \\
  &\quad \approx -(\frac{E}{\sqrt{\Delta_+^2-E^2}}c_+ F_+ + \frac{E}{\sqrt{\Delta_-^2-E^2}}c_- F_-) - (s_+ F_+ + s_- F_-), \\
  &\tilde{B}_{-} \equiv \int_0^{2\pi} \frac{d\varphi_k}{2\pi} ie^{i\varphi_k} B_{-} \\
  &\quad \approx (\frac{E}{\sqrt{\Delta_+^2-E^2}}s_+ F_+ - \frac{E}{\sqrt{\Delta_-^2-E^2}}s_- F_-) - (c_+ F_+ - c_- F_-), \\
  &\tilde{C}_{+} \equiv \int_0^{2\pi} \frac{d\varphi_k}{2\pi} C_{+} \\
  &\quad \approx -\frac{\Delta_+}{\sqrt{\Delta_+^2-E^2}} c_+ F_+ + \frac{\Delta_-}{\sqrt{\Delta_-^2-E^2}} c_- F_- , \\
  &\tilde{C}_{-} \equiv \int_0^{2\pi} \frac{d\varphi_k}{2\pi} ie^{i\varphi_k} C_{-} \\
  &\quad \approx \frac{\Delta_+}{\sqrt{\Delta_+^2-E^2}} s_+ F_+ - \frac{\Delta_-}{\sqrt{\Delta_-^2-E^2}} s_- F_-,
\end{align}
with
\begin{align}
  & c_b \equiv \cos[k_F^{(b)}r-\pi/4],\quad s_b \equiv \sin[k_F^{(b)}r-\pi/4], \\
  & F_b \equiv \frac{\sqrt{2\pi}\rho_b \,e^{-\sqrt{\Delta_b^2-E^2}r/v_F}}{\sqrt{k_F^{(b)}r}} = \sqrt{k_F^{(b)}}\,\frac{e^{-\sqrt{\Delta_b^2-E^2}r/v_F}}{\sqrt{2\pi r}v_F}.
\end{align}

Now we assume $\Delta_+ = \Delta_- = \Delta$, focusing on pure singlet superconductivity, and ignore $O(E^2/\Delta^2)$ terms (the deep-dilute-Shiba-state approximation). Then
\begin{align}
  & G_0(E,\bm{r}) \approx
  \frac{E}{\Delta}\tau_0\otimes\tilde{\alpha}(\bm{r})
  + \tau_x\otimes\tilde{\alpha}(\bm{r}) + \tau_z\otimes\tilde{\beta}(\bm{r}), \\
  & \tilde{\alpha}(\bm{r}) = -\frac{1}{2}(c_+F_+ + c_-F_-) - \frac{i}{2}(s_+ F_+ - s_- F_-)(\sigma_y\cos\varphi_r - \sigma_x\sin\varphi_r), \\
  & \tilde{\beta}(\bm{r}) = \frac{1}{2}(s_+F_+ + s_-F_-) - \frac{i}{2}(c_+ F_+ - c_- F_-)(\sigma_y\cos\varphi_r - \sigma_x\sin\varphi_r), \\
  & F_b \approx \sqrt{k_F^{(b)}}\,\frac{e^{-r/\xi}}{\sqrt{2\pi r}v_F} \qquad
  (\xi \equiv v_F/\Delta).
\end{align}

For an FM Shiba lattice, assume the magnetization is along $\hat{z}$, we have [cf. Eq.~\eqref{eq:G0Vi}],
\begin{align}
  \sum_{\bm{r}'} G_0(E,\bm{r}-\bm{r}')(-J\sigma_z\otimes\tau_0) \psi({\bm{r}'}) = \psi({\bm{r}}),
\end{align}
or,
\begin{align}
  &\frac{E}{\Delta} \sum_{\bm{r}'} M_1(\bm{r}-\bm{r}')\psi({\bm{r}'}) = \sum_{\bm{r}'} M_0(\bm{r}-\bm{r}')\psi({\bm{r}'}), \label{eq:preSchEq0} \\
  &M_1(\delta\bm{r}=0) = \left(\frac{1}{2}\,\pi J \sum_{b} \rho_b\right) \tau_0\otimes\sigma_z, \\
  &M_1(\delta\bm{r}\ne 0) = \left(\frac{1}{2}\, J\right)\tau_0\otimes[\tilde{c}_+\sigma_z - \tilde{s}_-(\sigma_x\cos\varphi_r + \sigma_y\sin\varphi_r)], \\
  &M_0(\delta\bm{r}=0) = \mathds{1} - \left(\frac{1}{2}\,\pi J \sum_{b} \rho_b\right) \tau_x\otimes\sigma_z, \\
  &M_0(\delta\bm{r}\ne 0) = \left(\frac{1}{2}\, J\right) \bigl\{ -\tau_x\otimes[\tilde{c}_+\sigma_z - \tilde{s}_-(\sigma_x\cos\varphi_r + \sigma_y\sin\varphi_r)] \nonumber \\
  &\hspace{40mm}+ \tau_z\otimes[\tilde{s}_+\sigma_z + \tilde{c}_-(\sigma_x\cos\varphi_r + \sigma_y\sin\varphi_r)] \bigr\},
\end{align}
where
\begin{align}
  &\rho_{\pm} \equiv k_F^{(\pm)}/2\pi v_F, \\
  &\tilde{c}_{\pm} \equiv c_+F_+ \pm c_-F_- = \frac{e^{-r/\xi}}{\sqrt{2\pi r}v_F} \left\{\sqrt{k_F^{(+)}}\cos[k_F^{(+)}r-\pi/4] \pm \sqrt{k_F^{(-)}}\cos[k_F^{(-)}r-\pi/4]\right\}, \\
  &\tilde{s}_{\pm} \equiv s_+F_+ \pm s_-F_- = \frac{e^{-r/\xi}}{\sqrt{2\pi r}v_F} \left\{\sqrt{k_F^{(+)}}\sin[k_F^{(+)}r-\pi/4] \pm \sqrt{k_F^{(-)}}\sin[k_F^{(-)}r-\pi/4]\right\}.
\end{align}

In the limit where the Shiba impurities are ultimately dilute, Eq.~\eqref{eq:preSchEq0} becomes
\begin{align}
  &\frac{E}{\Delta} \psi({\bm{r}}) = \left(\frac{1}{\eta} \tau_0\otimes\sigma_z - \tau_x\otimes\sigma_0\right) \psi({\bm{r}}), \\
  &\eta \equiv \frac{1}{2}\,\pi J \sum_{b} \rho_b = \frac{J}{2v_F}\,\sqrt{1+\frac{\alpha^2}{4}}.
\end{align}
Since we have already assumed the deep Shiba state limit, where $\eta\sim 1$, we find the two low energy states
\begin{align}
  &E_\pm = \pm\Delta\left(\frac{1}{\eta} - 1\right), \quad
  \psi_+ = \frac{1}{\sqrt{2}}
  \begin{pmatrix}
    1 \\ 0 \\ 1 \\ 0
  \end{pmatrix}, \quad
  \psi_- = \frac{1}{\sqrt{2}}
  \begin{pmatrix}
    0 \\ 1 \\ 0 \\ -1
  \end{pmatrix}.
\end{align}
Projected into the subspace spanned by these two states, Eq.~\eqref{eq:preSchEq0} becomes
\begin{align}
  &\frac{E}{\Delta} \sum_{\bm{r}'} \tilde{M}_1(\bm{r}-\bm{r}') \tilde{\psi}({\bm{r}'}) = \sum_{\bm{r}'} \tilde{M}_0(\bm{r}-\bm{r}') \tilde{\psi}({\bm{r}'}), \label{eq:preSchEq1} \\
  &\tilde{M}_1(\delta\bm{r}=0) = \eta \tilde{\sigma}_z, \\
  &\tilde{M}_1(\delta\bm{r}\ne 0) = (\frac{1}{2}\, J)\tilde{c}_+\tilde{\sigma}_z, \\
  &\tilde{M}_0(\delta\bm{r}=0) = (1 - \eta) \tilde{\sigma}_0, \\
  &\tilde{M}_0(\delta\bm{r}\ne 0) = (\frac{1}{2}\, J) \bigl[ -\tilde{c}_+\tilde{\sigma}_0
  +\tilde{c}_-(\tilde{\sigma}_x\cos\varphi_r + \tilde{\sigma}_y\sin\varphi_r) \bigr],
\end{align}
where
\begin{align}
  \tilde{M}_i \equiv (\psi_+,\psi_-)^\dag M_i (\psi_+,\psi_-)
  \qquad (i=0,1),
\end{align}
$\tilde{\psi}$ and $\tilde{\bm{\sigma}}$ (tildes will be omitted hereafter) are wavefunctions and Pauli matrices under the basis $(\psi_+, \psi_-)$. Fourier transforming Eq.~\eqref{eq:preSchEq1} we obtain
\begin{align}
  &H_{\text{eff}}(\bm{q}) \psi(\bm{q}) = \frac{E}{\Delta} \psi(\bm{q}), \\
  &H_{\text{eff}}(\bm{q}) \equiv \left[\sum_{\bm{r}} e^{-i\bm{q}\cdot\bm{r}}\tilde{M}_1(\bm{r})\right]^{-1} \left[\sum_{\bm{r}} e^{-i\bm{q}\cdot\bm{r}}\tilde{M}_0(\bm{r})\right].
\end{align}
Or,
\begin{align}
  &H_{\text{eff}}(\bm{q}) = \left[\eta + d_0(\bm{q})\right]^{-1} \left\{[1 - \eta - d_0(\bm{q})] {\sigma}_z + d_1(\bm{q}) {\sigma}_x + d_2(\bm{q}) {\sigma}_y\right\}, \label{eq:Heff} \\
  & d_0(\bm{q}) \equiv \left(\frac{1}{2}\, J\right) \sum_{\bm{r}\ne 0} e^{-i\bm{q}\cdot\bm{r}}\tilde{c}_+(r), \\
  & d_1(\bm{q}) \equiv -\left(\frac{i}{2}\, J\right) \sum_{\bm{r}\ne 0} e^{-i\bm{q}\cdot\bm{r}}\tilde{c}_-(r) \sin\varphi_r, \\
  & d_2(\bm{q}) \equiv \left(\frac{i}{2}\, J\right) \sum_{\bm{r}\ne 0} e^{-i\bm{q}\cdot\bm{r}}\tilde{c}_-(r) \cos\varphi_r,
\end{align}
Note that $d_1$ and $d_2$ are defined with slight differences in the main text in order to make a compact expression.

We can further construct a Majorana basis by using the fact that $\psi_+$ and $\psi_-$ are particle-hole images of each other:
\begin{align}
  \psi_1^{(M)} = \frac{1}{\sqrt{2}}(\psi_++\psi_-) =
  \frac{1}{2}
  \begin{pmatrix}
    1 \\ 1 \\ 1 \\ -1
  \end{pmatrix}, \;
  \psi_2^{(M)} = \frac{i}{\sqrt{2}}(\psi_+-\psi_-) =
  \frac{i}{2}
  \begin{pmatrix}
    1 \\ -1 \\ 1 \\ 1
  \end{pmatrix}.
\end{align}
Under this Majorana basis, the effective Hamiltonian is given by
\begin{align}
  &H_{\text{eff}}^{(M)}(\bm{q}) = -\left[\eta + d_0(\bm{q})\right]^{-1} \left\{[1 - \eta - d_0(\bm{q})] {\sigma}_y + d_2(\bm{q}) {\sigma}_x - d_1(\bm{q}) {\sigma}_z\right\}.
\end{align}
Clearly, when $\bm{q}$ is one of the inversion symmetric momenta [ISMs, namely, $(0,0)$, $(0,\pi)$, $(\pi,0)$, or $(\pi,\pi)$], $ e^{-i\bm{q}\cdot\bm{r}} =  e^{i\bm{q}\cdot\bm{r}}$, therefore $d_1(\bm{q}) = d_2(\bm{q}) = 0$; the Pfaffian function is given by
\begin{align}
  P(\bm{q}\in\text{ISMs}) &= 1- \frac{1}{\eta + d_0(\bm{q})} \\
  &\simeq d_0(\bm{q}) \qquad (\eta\simeq 1).
  \label{eq: pfaffian}
\end{align}

Now return to the effective Hamiltonian Eq.~\eqref{eq:Heff}. If spin-orbit coupling is absent, then $\forall \bm{q}: \; d_1(\bm{q}) = d_2(\bm{q}) = 0$. This immediately implies that if there is any sign change in $P(\bm{q})$ among the ISMs, the system is gapless with line nodes. { Rashba spin-orbit coupling potentially lifts these degeneracies. 
The sign of the Pfaffian formula~\eqref{eq: pfaffian}, multiplied over both $C_4$ symmetric momenta is equal to the parity of the Chern number. We have evaluated the expression numerically, with the result given in the main text. }

\begin{figure}
  \centering
  \includegraphics[width=0.7\textwidth]{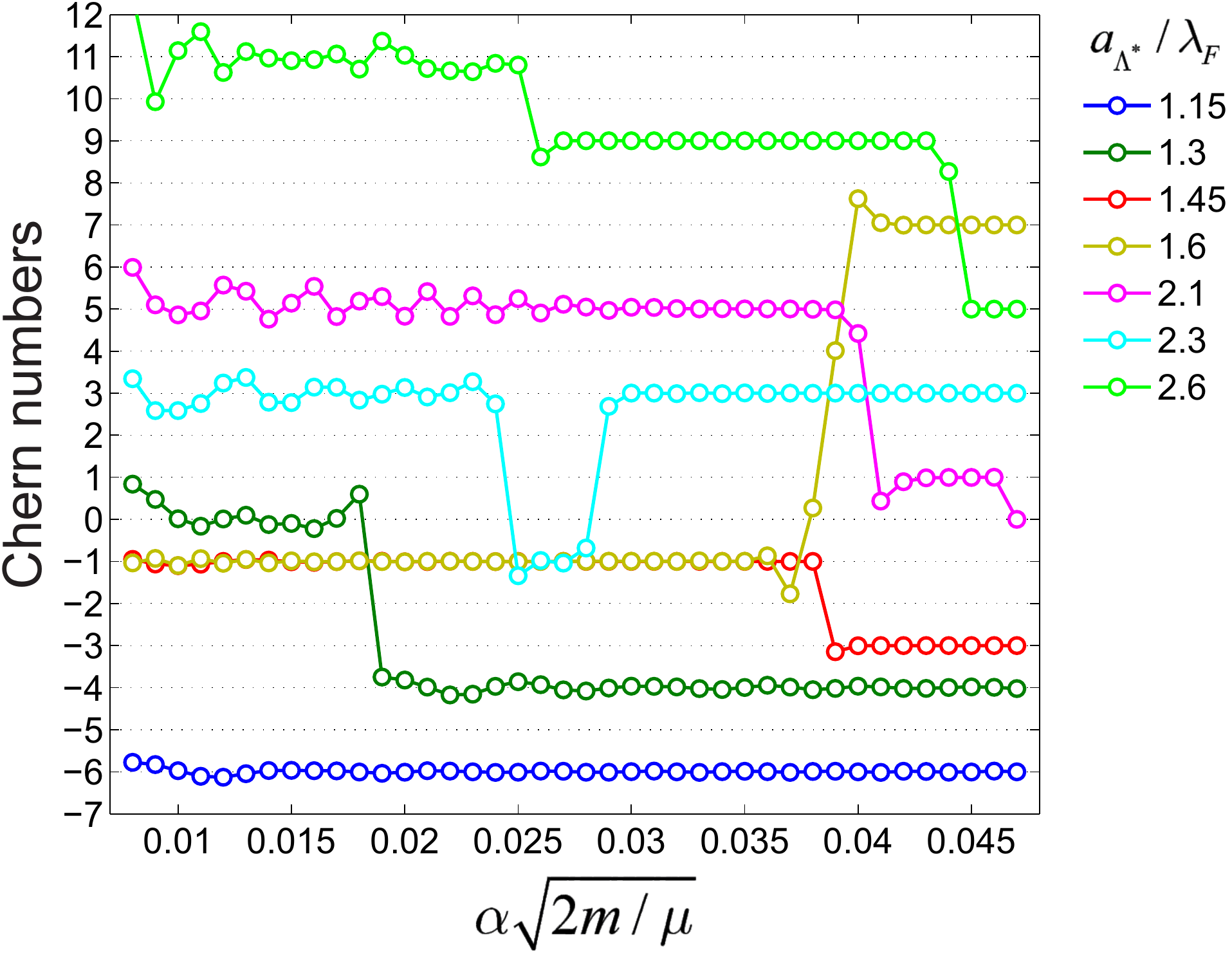}
  \caption{Chern numbers computed straightforwardly from Hamiltonian~\eqref{eq:Heff} as a function of spin-orbit coupling strength, for several values of lattice spacing (in units of Fermi wavelength $\lambda_F$) of Shiba impurities. The ill quantization of some of the Chern numbers is due to numerical errors. Note that the symbols here are dimensional to be consistent with the main text.}
  \label{fig:shiba2deff_cns}
\end{figure}

In addition we have evaluated the Chern numbers through numerical integrations by using Hamiltonian~\eqref{eq:Heff} straightforwardly. Some representative results are shown in Fig.~\ref{fig:shiba2deff_cns}. We confirm the existence of high Chern numbers associated with this effective Hamiltonian, which complements the phase diagram in terms of the parity of the Chern numbers presented in Fig.~3 of the main text.

\section{Two-dimensional Shiba lattice with helical magnetic order}
\label{sec: helix}

In this section we focus on a different type of magnetic order: we assume a two-dimensional helical pattern for the magnetic moments. For simplicity we neglect the detailed spatial structure of the Shiba states (which is crucial in the previous section) and investigate the following tight-binding Hamiltonian
\begin{align}
  H = \sum\limits_n\Biggl\{
  &c^\dagger_n(\bm{B}_n\cdot\bm{\sigma}-\mu)c_n
  + [\Delta_{n} c_n^T (\mathrm{i}\sigma_y) c_n + h.c.] +\sum\limits_{\delta=\pm1_x,\pm1_y}
  t_{n\delta} \, c^\dagger_n  c_{n+\delta}\Biggr\},
\end{align}
where $n$ is a 2D index for the tight-binding sites, $c_n = (c_{n\uparrow}, c_{n\downarrow})^T$ is the vector of electron annihilation operators at site $n$, $\bm{B}_n$ stands for the local magnetic moment coupled to the superconductor electron spin, $\Delta_{n}$ stands for the local pairing potential, and $t_{n\delta}$ stands for spin-independent nearest-neighbor hoppings.

By changing to a rotating basis $g_n  = U_n^\dag c_n$ with $U_n$ defined by
\begin{align}
  U_n^\dagger (\bm{B}_n\cdot\bm{\sigma}) U_n = B_n\sigma_z, \quad
  U_n^\dagger U_n = \sigma_0,\quad \text{Det}\,U_n = 1,
\end{align}
the above Hamiltonian becomes
\begin{align}\label{eq:ham_rot_bas}
  \tilde{H} = \sum\limits_n\Biggl\{
  &g^\dagger_n(B_n\sigma_z-\mu)g_n
  + [\Delta_{n} g_n^T (\mathrm{i}\sigma_y) g_n + h.c.] +\sum\limits_{\delta=\pm1_x,\pm1_y}
  t_{n\delta} \, g^\dagger_n \Omega_{n\delta}  g_{n+\delta}\Biggr\},
\end{align}
where
\begin{align}
  \Omega_{n\delta} = U_{n}^\dagger U_{n+\delta},\quad \text{Det}\,\Omega_{n\delta} = 1 .
\end{align}
To further simplify the problem, we set $B_n = B$, $\Delta_n=\Delta_n^*=\Delta$, $t_{n1_x}=t_{n1_x}^*=t_x$, $t_{n1_y}=t_{n1_y}^*=t_y$ for all $n$.

We also assume a certain periodical pattern for the magnetic moments $\hat{B}_n$. One simplest, nevertheless quite general, choice is to let
\begin{align}
  &\Omega_{(n_x,0),\delta=1_x} = \Omega_{x} = \exp[i(\rho_x/2)\hat{r}_x\cdot\bm{\sigma}], \\
  &\Omega_{(n_x,n_y),\delta=1_y} = \Omega_{y} = \exp[i(\rho_y/2)\hat{r}_y\cdot\bm{\sigma}],
\end{align}
for all $n_x$ and $n_y$. Here $\rho_{x,y}$ stand for rotation angles between neighboring sites and $\hat{r}_{x,y}$ stand for rotation axes. The rest of the $\Omega$-matrices are hereby determined by a closed-path formula:
\begin{align}\label{eq:Omega_closed_path}
  \Omega_{(n_x,n_y),\delta=1_x} = \Omega^{(n_y)}_{x} = \Omega_{y}^{-n_y}\Omega_{x}\Omega_{y}^{n_y} = \exp[i(\rho_x/2)\Omega_{y}^{-n_y}(\hat{r}_x\cdot\bm{\sigma})\Omega_{y}^{n_y}].
\end{align}
If $\Omega_{x}$ and $\Omega_{y}$ do not commute and $\rho_y = 2\pi p/q$ [which implies $\Omega_{y}^q = (-\mathds{1})^p$] with $p$ and $q$ relatively prime integers, then $\Omega^{(n_y)}_{x}=\Omega^{(n_y\mod q)}_{x}$; each unit cell consists of $q$ sites. The Bloch Hamiltonian in the Nambu basis is given by
\begin{subequations}\label{eq:ham2dk}
\begin{align}
  &H_k =
  \bm{g}_k^\dagger
  \begin{pmatrix}
    h(B,\bm{k}) & \Delta \\
    \Delta & -h(-B,\bm{k})
  \end{pmatrix}
  \bm{g}_k, \\
  &\bm{g}_k^\dagger = (g_{0,k\uparrow}^\dagger, g_{0,k\downarrow}^\dagger,\ldots, g_{q-1,k\uparrow}^\dagger, g_{q-1,k\downarrow}^\dagger, g_{0,-k\downarrow}, -g_{0,-k\uparrow},\ldots, g_{q-1,-k\downarrow}, -g_{q-1,-k\uparrow}), \\
  &h(B,\bm{k}) =
  \begin{pmatrix}
    h_{0}(B,k_x) & \tau(k_y) & 0 & \tau(k_y)^{\dagger} \\
    \tau(k_y)^{\dagger} & h_{1}(B,k_x) & \tau(k_y) & 0 \\
    0 & \ddots & \ddots & \ddots \\
    \tau(k_y) & 0 & \tau(k_y)^{\dagger} & h_{q-1}(B,k_x)	
  \end{pmatrix}, \\
  &h_{n_y}(B,k_x)={B}{\sigma}_z-\mu+t_x(\Omega^{(n_y)}_x e^{ik_x}+h.c.) , \\
  &\tau(k_y)=t_y\Omega_y e^{ik_y/q}.
\end{align}
\end{subequations}

The above model is similar, but in general not equivalent, to a model with constant $\bm{B}_n$ and spin-orbit coupling, such as the following one:
\begin{align}\label{eq:hso_comp}
  h_{SO}(B,\bm{k}) = \frac{\hbar^2 k^2}{2m} - \mu + {B}{\sigma}_z + k_x(\bm{\alpha}_x\cdot\bm{\sigma}) + k_y(\bm{\alpha}_y\cdot\bm{\sigma}),
\end{align}
where $\bm{\alpha}_x$ and $\bm{\alpha}_y$ are real-valued vectors. To see this, we write the Hamiltonian \eqref{eq:hso_comp} on a lattice. The hopping terms have the form
\begin{align}
  \langle n | h_{SO} | n+1_{l=x,y} \rangle = t + \frac{1}{2i}\bm{\alpha}_l\cdot\bm{\sigma}.
\end{align}
If we define
\begin{align}
  t_{l=x,y} = \sqrt{t^2 + |\bm{\alpha}_l/2|^2}, \quad \Omega_{l=x,y} &= (t + \frac{1}{2i}\bm{\alpha}_l\cdot\bm{\sigma})/t_l,
\end{align}
where $\Omega_{l=x,y}$ are unitary and $\text{Det}\,\Omega_{l}=1$, this Hamiltonian formally resembles Hamiltonian \eqref{eq:ham_rot_bas} except that, importantly, the $\Omega$'s  such defined do not necessarily satisfy a closed-path equation as Eq.~\eqref{eq:Omega_closed_path}.

\subsection{Commuting helix rotations}\label{ssec:com}
\begin{figure}
  \centering
  \includegraphics[width=0.9\textwidth]{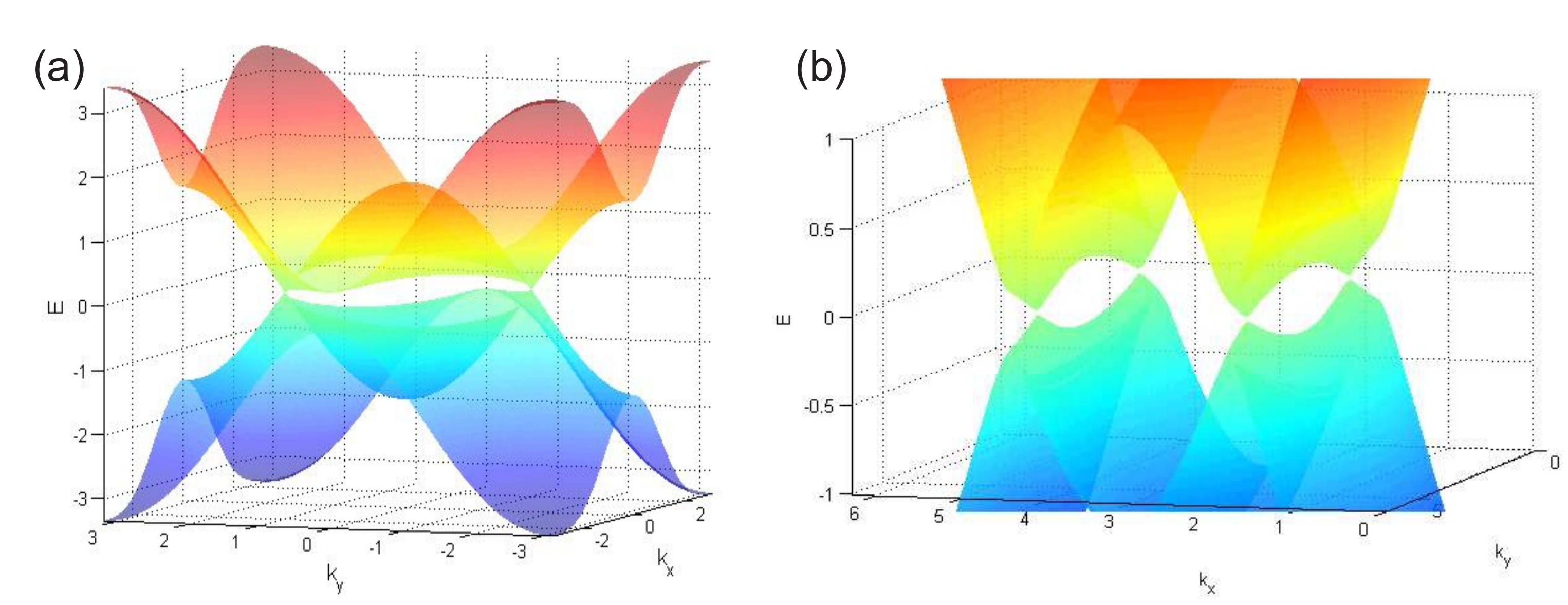}
  \caption{Examples of nodal superconductor band structures with (a) two Fermi points (see Sec.~\ref{ssec:com}), (b) four Fermi points (see Sec.~\ref{ssec:noncom}).}
  \label{fig:nodal_sc}
\end{figure}
Obviously, $\Omega_x$ and $\Omega_y$ commute when $\hat{r}_x = \hat{r}_y = \hat{r}$. In this case
\begin{align}
  h(B,\bm{k}) = {B}{\sigma}_z-\mu+t_x(\Omega_x e^{ik_x}+h.c.)+t_y(\Omega_y e^{ik_y}+h.c.).
\end{align}
Obviously we can rotate the basis again such that $\hat{r}=(\sin\theta,0,\cos\theta)$. Then
\begin{subequations}
\begin{align}
  &h(B,\bm{k}) =
  \begin{pmatrix}
    B-v-m & -u \\
    -u & -(B-v)-m
  \end{pmatrix}, \\
  &m = \mu - 2t_x\cos\frac{\rho_x}{2}\cos{k_x} - 2t_y\cos\frac{\rho_y}{2}\cos{k_y}, \\
  &u = (2t_x\sin\frac{\rho_x}{2}\sin{k_x} + 2t_y\sin\frac{\rho_y}{2}\sin{k_y})\sin\theta, \\
  &v = (2t_x\sin\frac{\rho_x}{2}\sin{k_x} + 2t_y\sin\frac{\rho_y}{2}\sin{k_y})\cos\theta.
\end{align}
\end{subequations}

Let us assume $\theta=\pi/2$, the spectrum of the Nambu Hamiltonian is given by
\begin{align}
  E &= \pm\sqrt{B^2+\Delta^2+m^2+u^2
  \pm 2\sqrt{B^2\Delta^2+B^2m^2+m^2u^2}}.
\end{align}
The conditions for $E=0$ are
\begin{align}
  &2t_x\sin\frac{\rho_x}{2}\sin{k_x} + 2t_y\sin\frac{\rho_y}{2}\sin{k_y} = 0, \\
  &\mu=2t_x\cos\frac{\rho_x}{2}\cos{k_x} + 2t_y\cos\frac{\rho_y}{2}\cos{k_y} \pm\sqrt{B^2-\Delta^2}.
\end{align}
Considering $B\gg\max(\Delta,t)$ and $\mu>0$, the spectrum is clearly gapped when $|\mu - \sqrt{B^2-\Delta^2}| > 2t_x|\cos\frac{\rho_x}{2}| + 2t_y|\cos\frac{\rho_y}{2}|$; otherwise the spectrum is gapless with two Fermi points $\pm(k_{x0},k_{y0})$ [see e.g. Fig.~\ref{fig:nodal_sc}(a)] that merge when the equal sign is taken.

\subsection{Non-commuting helix rotations}\label{ssec:noncom}
Now we consider the case when $\Omega_x$ and $\Omega_y$ do not commute. The most important feature of the band structure is that the subbands created by the multiple sites of a unit cell are \textit{not} gapped. To see this, let us look at the Hamiltonian \eqref{eq:ham2dk} at $k_x = 0$ or $\pi$, where $e^{ik_x}=\pm1$, and
\begin{align}
  \Omega^{(m)}_x e^{ik_x}+h.c. = \pm(\Omega^{(m)}_x+h.c.) = \pm 2\cos(\rho_x/2).
\end{align}
This implies that $h_m(B,k_x)$ has no dependence on $m$ at $k_x = 0$ or $\pi$; the unit cell is essentially of a single site. Explicitly, the Hamiltonian becomes
\begin{align}
  &H_{k_x=0,\pi}(k_y) = \sum\limits_{m=0}^{q-1}
  \tilde{\bm{g}}_{mk_y}^\dagger
  \begin{pmatrix}
    \tilde{h}_m(B,k_y) & \Delta \\
    \Delta & -\tilde{h}_m(-B,k_y)
  \end{pmatrix}
  \tilde{\bm{g}}_{mk_y}, \\
  &\tilde{\bm{g}}_{mk}^\dagger = (\tilde{g}_{m,k\uparrow}^\dagger, \tilde{g}_{m,k\downarrow}^\dagger, \tilde{g}_{m,-k\downarrow}, -\tilde{g}_{m,-k\uparrow}),\quad \tilde{g}_{m,k\sigma} = \sum\limits_{m'=0}^{q-1} \frac{1}{\sqrt{q}} e^{-i \frac{2\pi mm'}{q}} g_{m',k\sigma}, \\
  &\tilde{h}_m(B,k_y) ={B}{\sigma}_z-\mu \pm 2t_x\cos(\rho_x/2) + \tau(2\pi m+k_y) + \tau(2\pi m+k_y)^\dagger.
\end{align}
The crossings of the subbands do not necessarily occur at $k_y=0$ or $\pi$ unless there is an inversion symmetry. One example is that if $\Omega_{y} = \exp[i(\rho_y/2)(\sigma_x\cos\varphi_y+\sigma_y\sin\varphi_y)]$, then $\sigma_z\Omega_{y}\sigma_z = \Omega_{y}^\dagger$, and $\sigma_z\tilde{h}_m(B,k_y)\sigma_z = \tilde{h}_{q-m}(B,-k_y)$. In other words, if the rotation axis $\hat{r}_y$ lies in the $x$-$y$ plane, then the crossings of the subbands occur at $k_y=0$ or $\pi$.

As a special case, we set $\rho_y=\pi$, $\hat{r}_y=(1,0,0)$, $\hat{r}_x=(\cos\varphi_x,\sin\varphi_x,0)$.
In this case, each unit cell consists of two sites, $\Omega_{y}=i\sigma_x$, $\Omega_{y}^{-1}\Omega_{x}\Omega_{y}=
  \Omega_{x}^T$, and (we also set $t_x=t_y=t$)
\begin{align}
  &h(B,\bm{k}) =
  \begin{pmatrix}
    {B}{\sigma}_z-\mu+t(\Omega_x e^{ik_x}+h.c.) & -2t\sigma_x \sin{k_y/2} \\
    -2t\sigma_x \sin{k_y/2} & {B}{\sigma}_z-\mu+t(\Omega_{x}^T e^{ik_x}+h.c.)	
  \end{pmatrix}.
\end{align}
The spectrum of this Hamiltonian can be solved to be
\begin{align}\label{eq:sp2d}
  E &= \pm\sqrt{B^2+\Delta^2+m_k^2+d_{k\pm}^2
  \pm 2\sqrt{B^2\Delta^2+B^2m_k^2+m_k^2d_{k\pm}^2}}, \\
  m_k &= \mu-2t\cos\frac{\rho_x}{2}\cos{k_x}, \\
  d_{k\pm} &= 2t(\sin\frac{\rho_x}{2}\sin{k_x} \pm \sin\frac{k_y}{2}).
\end{align}
The conditions for $E=0$ are
\begin{align}
  &\sin\frac{\rho_x}{2}\sin{k_x} \pm \sin\frac{k_y}{2} = 0, \\
  &\mu=2t\cos\frac{\rho_x}{2}\cos{k_x}\pm\sqrt{B^2-\Delta^2}.
\end{align}
Considering $B\gg\max(\Delta,t)$ and $\mu>0$, the spectrum is gapped only when $\mu>|2t\cos\frac{\rho_x}{2}|+\sqrt{B^2-\Delta^2}$ or $\mu<-|2t\cos\frac{\rho_x}{2}|+\sqrt{B^2-\Delta^2}$. When $|\mu - \sqrt{B^2-\Delta^2}| < |2t\cos\frac{\rho_x}{2}|$, there are always four Fermi points [see e.g. Fig.~\ref{fig:nodal_sc}(b)].

\begin{figure}
  \centering
  \includegraphics[width=\textwidth]{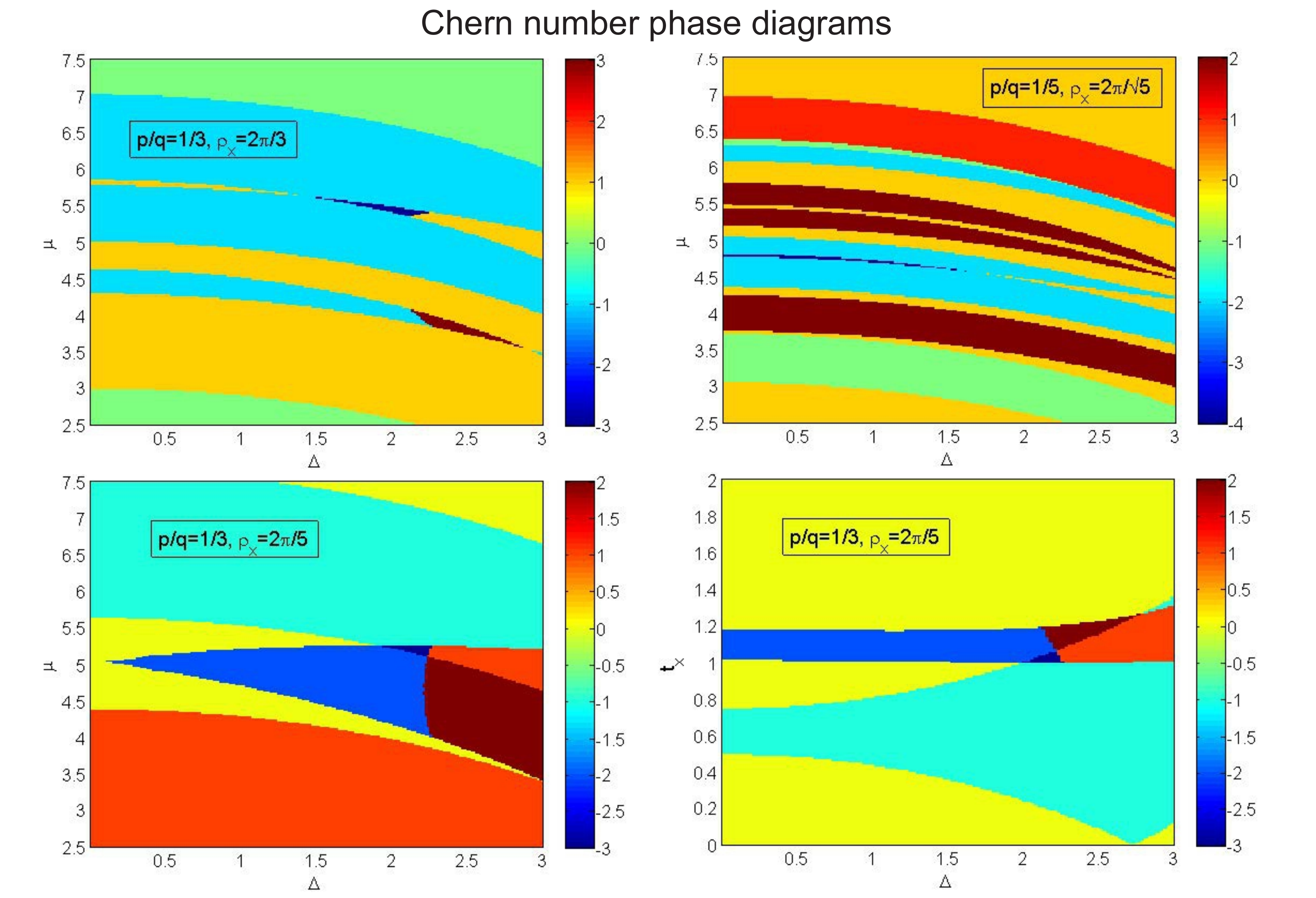}
  \caption{Examples of the Chern number phase diagrams with different helical patterns.}
  \label{fig:helix_pds}
\end{figure}
\begin{figure}
  \centering
  \includegraphics[width=0.9\textwidth]{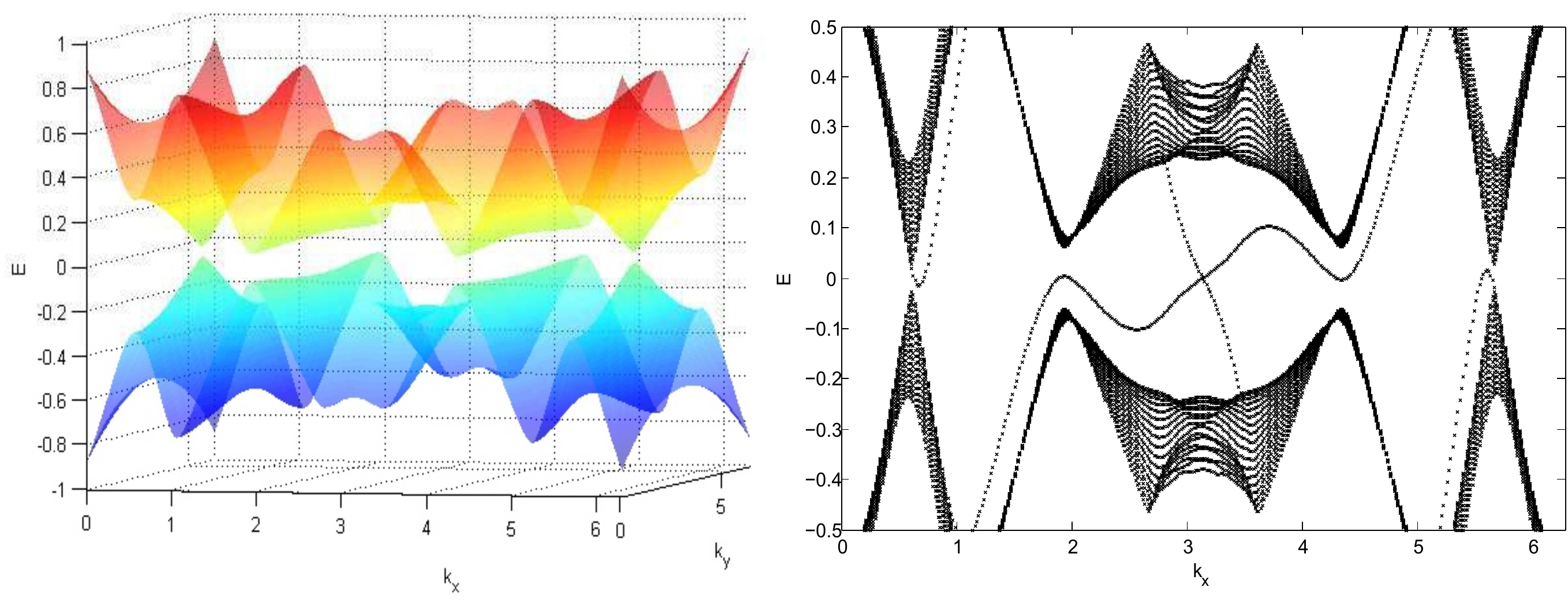}
  \caption{Low-energy bulk bands (left panel) and dispersion relations for an open-boundary strip (right panel) with $p/q=1/3$, $\rho_x=2\pi/3$, $\mu=5.5$, $\Delta = 0.9$. The Chern number is $-1$ in this example.}
  \label{fig:q3thx3}
\end{figure}

With more general helical patterns, full superconducting gaps can be opened with nontrivial topological properties described by Chern numbers \cite{nakosai_2013}. The Chern number, as well as the spectrum for an open-boundary sample, can be computed numerically. Unless otherwise stated, we set $B= 5$, $t_x=t_y=1$, $\hat{r}_y=(1,0,0)$ and $\hat{r}_x=(0,-1,0)$ in our computations. We shown in Fig.~\ref{fig:helix_pds} several examples of the phase diagrams in terms of the Chern numbers with different helical patterns. In Fig.~\ref{fig:q3thx3} we show in addition an example of bulk and edge spectrum for a fixed combination of $\mu$ and $\Delta$.

\section{Details of density functional theory analyses}

We performed electronic structure calculations within the density functional theory (DFT) formalism as implemented in the Vienna ab initio simulation package \cite{Kresse_1993}. We used the all-electron projector augmented wave (PAW) \cite{Blochl_1994, Kresse_1999} basis sets with the generalized gradient approximation of Perdew, Burke and Ernzerhof \cite{Perdew_1996} to the exchange correlation potential. The Hamiltonian contains scalar relativistic corrections, and the spin-orbit coupling was taken into account by the second variation method \cite{Koelling_1977}.

\begin{figure}
  \centering
  \includegraphics[width=0.6\textwidth]{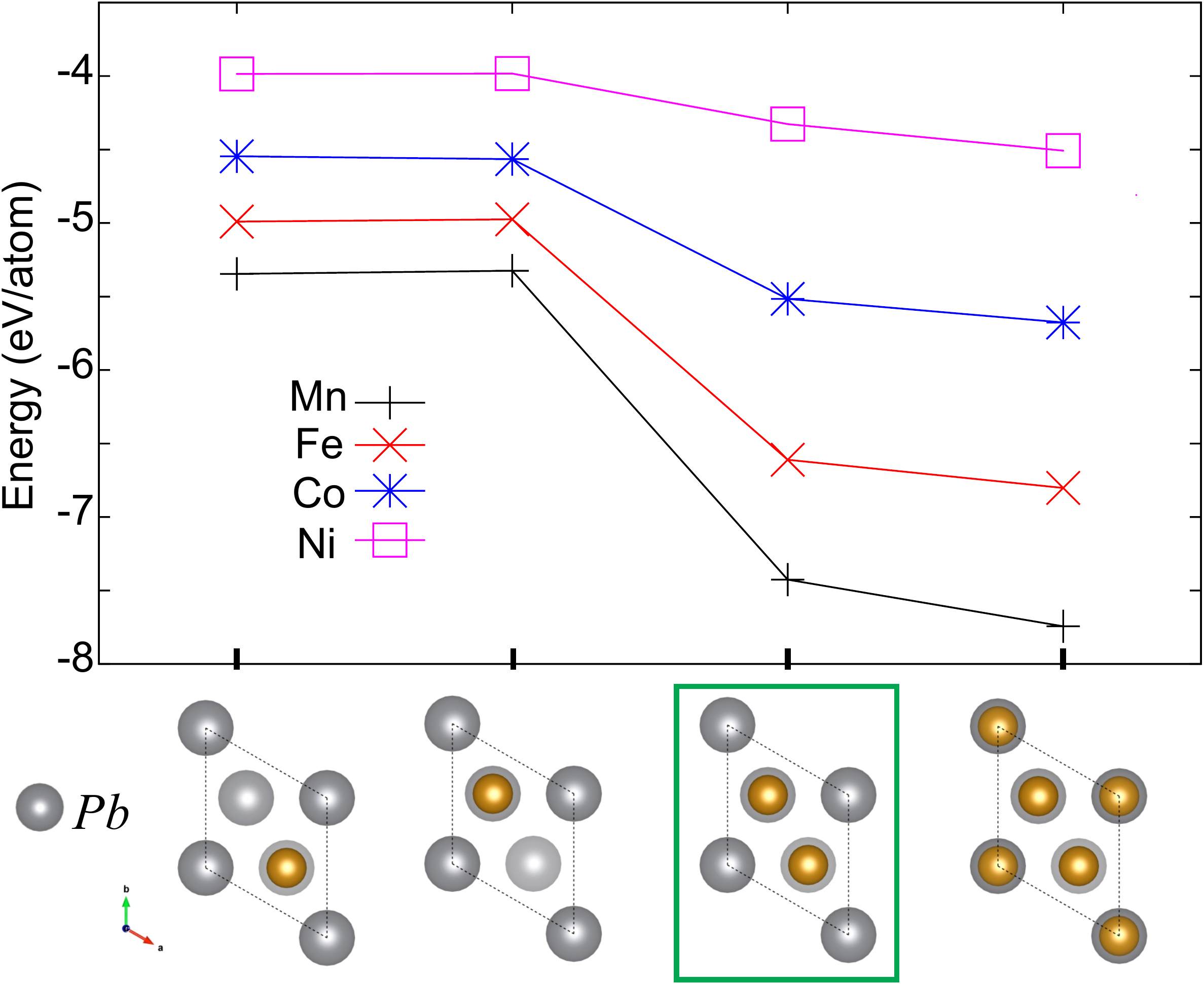}
  \caption{Relaxation energy of different lattice configurations for different transition metal adatoms. The configuration adopted in our simulations is highlighted.}
  \label{fig:lattice_relaxation}
\end{figure}
\begin{figure}
  \centering
  \includegraphics[width=0.6\textwidth]{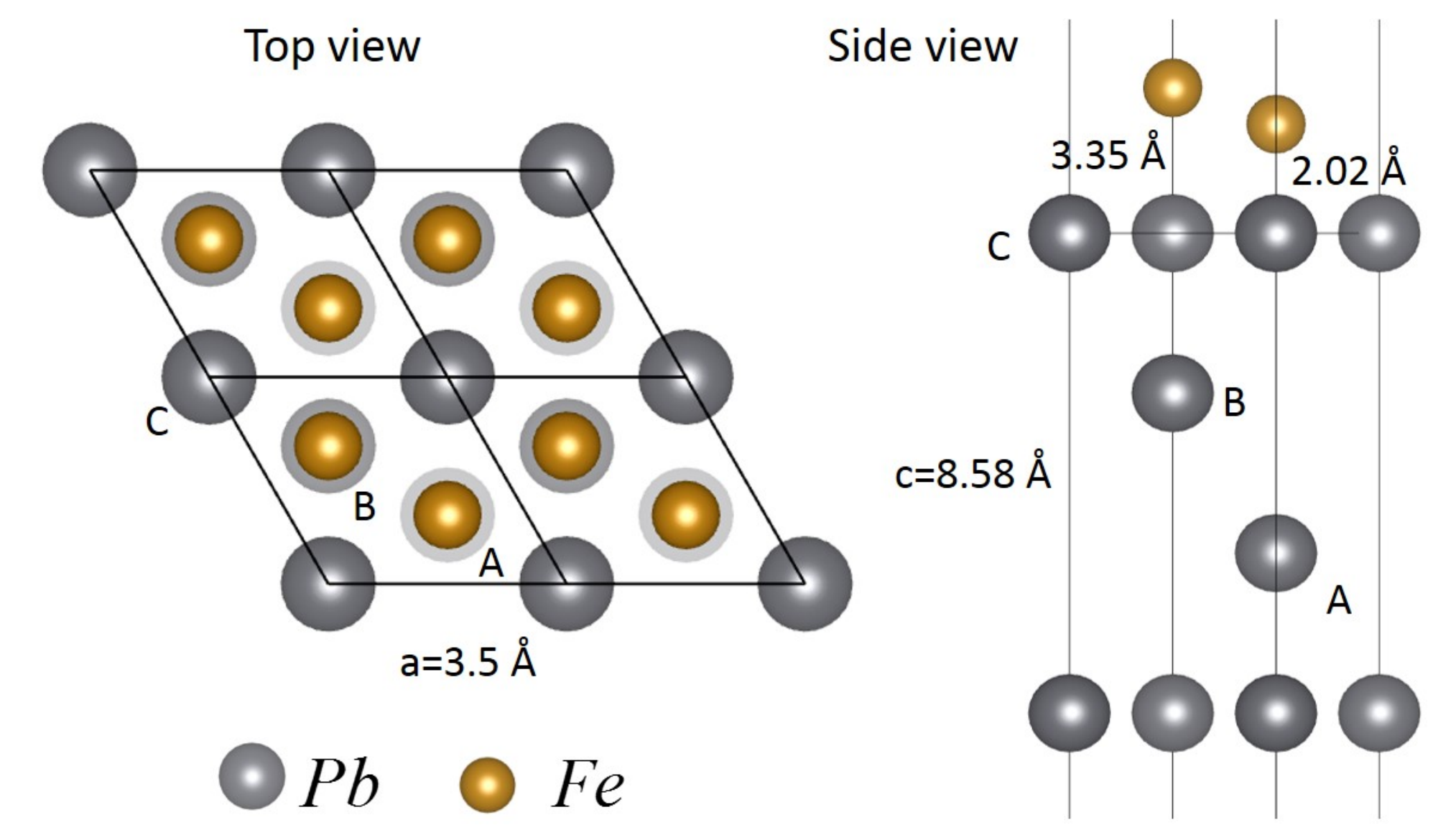}
  \caption{Details of the atomic configuration adopted in our simulations. This configuration has a notable $C3$ symmetry.}
  \label{fig:dft_lattice}
\end{figure}

In this work, we chose the host superconductor to be Pb thin film with a (111) surface, and consider different transition metal adatoms. We started by finding the stable configurations of the adatoms on top of the Pb surface. To this end we have compared the relaxation energy (per atom) for an extensive collection of possible configurations, four of which are shown in Fig.~\ref{fig:lattice_relaxation}. Based on this energetic consideration and for simplicity, we adopted the configuration with two adatoms per Pb unit cell (highlighted in Fig.~\ref{fig:lattice_relaxation}) in our following simulations. When the transition metal element is chosen to be Fe, the details of the configuration are shown in Fig.~\ref{fig:dft_lattice}. In this configuration, two species of Fe atoms form a buckled honeycomb structure which gains bonding energy due to the short nearest-neighbor distance.

To simulate the composite system and consider the effect of Pb, we used 6 layers of Pb atoms as substrate in the relaxation calculations with roughly 15 \AA $\ $vacuum space, taking into account spin-orbital coupling and the ferromagnetic alignment of the Fe moments. We performed DFT calculations with the stable configuration as shown in Fig.~\ref{fig:dft_lattice}. Based on the DFT calculations, we constructed the maximally localized Wannier functions \cite{Marzari_1997, Souza_2001} for Fe, and obtained a tight-binding model with a band structure that agrees well with the DFT result. We then used the tight-binding model and added a small $s$-wave superconducting pairing term to it. We computed the Chern numbers of the thus obtained Bogoliubov-deGennes Hamiltonian to be nonvanishing, as shown in Fig.~4 from the main text. For completeness, we present a few more Fermi surfaces with different values of $E_F$ in Fig.~\ref{fig:fermi_surfaces} to complement Fig.~4 from the main text.

\begin{figure}
  \centering
  \includegraphics[width=0.8\textwidth]{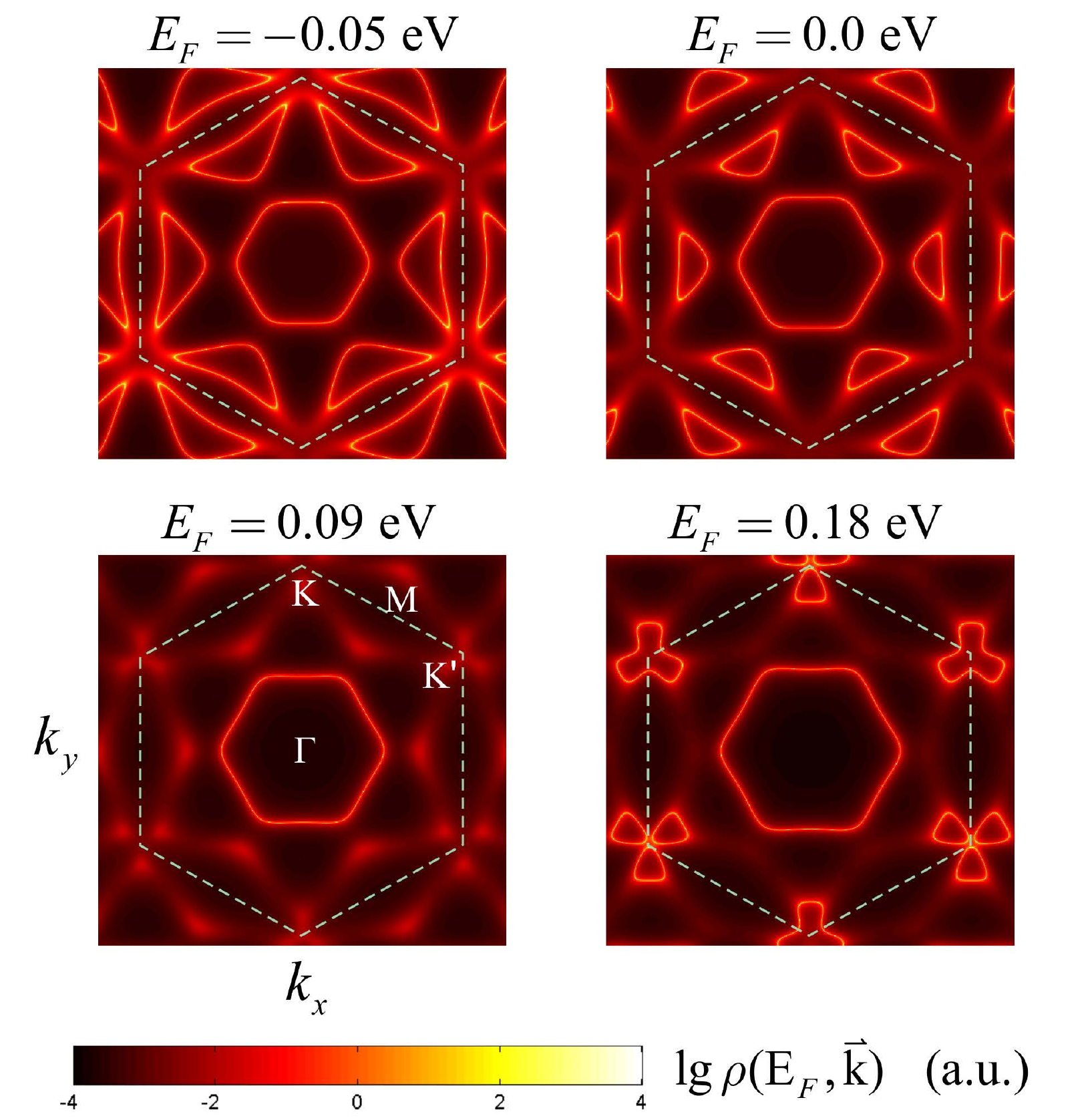}
  \caption{Fermi surfaces with four values of $E_F$ around 0. In the presence of a small (0.01eV) s-wave superconducting pairing potential, the Chern number corresponding to the superconducting phase with $E_F=0.18$ eV is 3, and the Chern numbers in the rest cases are all 1 (see main text Fig.~4).}
  \label{fig:fermi_surfaces}
\end{figure}

\end{widetext}

\end{document}